\DocumentMetadata{testphase=new-or-1}
\documentclass[fleqn]{aa}

\usepackage{newtxtext,newtxmath,dsfont}
\usepackage[T1]{fontenc}

\DeclareRobustCommand{\VAN}[3]{#2}
\let\VANthebibliography\thebibliography
\def\thebibliography{\DeclareRobustCommand{\VAN}[3]{##3}\VANthebibliography}

\providecommand{\tightlist}{}

\usepackage{graphicx}	%
\usepackage[pdfencoding=auto,psdextra]{hyperref}
\hypersetup{
    colorlinks=true,
    linkcolor=blue,
    filecolor=magenta,      
    urlcolor=blue,
    citecolor=blue
}
\usepackage{amsmath}	%
\usepackage{enumitem}
\usepackage{tikz}
\usetikzlibrary{positioning}
\usepackage{mathtools}
\usepackage{multirow}
\usepackage{bigdelim}
\usepackage{physics}
\usepackage{stfloats}
\usepackage[separate-uncertainty,separate-uncertainty-units=single]{siunitx}
\usepackage{booktabs}
\usepackage[normalem]{ulem}
\usepackage[noabbrev,capitalize]{cleveref}

\usepackage{xspace}
\newcommand*{\borg}{\textsc{borg}\xspace}
\newcommand{\Identity}{\mathbf{\mathds{1}}}

\makeatletter
\renewcommand*\aa@pageof{, page \thepage{} of \pageref*{LastPage}}
\makeatother

\DeclareSIUnit{\MSun}{\ensuremath{M_\odot}} 
\DeclareSIUnit{\Msun}{\ensuremath{M_\odot}} 

\newcommand{\LCDM}{$\Lambda$CDM\xspace}

\begin{document}

   \title{Constrained cosmological simulations of the Local Group using Bayesian hierarchical field-level inference}

   \author{Ewoud
          Wempe \inst{1}
          \and
          Guilhem Lavaux \inst{2}
          \and
          Simon D.M. White \inst{3}
          \and
          Amina Helmi \inst{1}
          \and
          Jens Jasche \inst{4}
          \and 
          Stephen Stopyra \inst{4}
          }

   \institute{Kapteyn Astronomical Institute, University of Groningen, P.O Box 800, 9700 AV Groningen, The Netherlands\\
              \email{ewoudwempe@gmail.com}
         \and
            CNRS \& Sorbonne Université, UMR 7095, Institut d’Astrophysique de Paris, 98 bis boulevard Arago, F-75014 Paris, France
        \and
            Max-Planck-Institut für Astrophysik, Karl-Schwarzschild-Straße 1, 85748 Garching, Germany
        \and 
            The Oskar Klein Centre, Department of Physics, Stockholm University, Albanova University Center, SE 106 91 Stockholm, Sweden
             }

   \date{Received X; accepted Y}

\abstract
{
We present a novel approach based on Bayesian field-level inference capable of resolving individual galaxies within the Local Group (LG), enabling detailed studies of its structure and formation via posterior simulations.
We extend the Bayesian Origin Reconstruction from Galaxies (\borg) algorithm with a multi-resolution approach, allowing us to reach smaller mass scales and apply observational constraints based on LG galaxies. Our updated data model simultaneously accounts for observations of mass tracers within the dark haloes of the Milky Way (MW) and M31, their observed separation and relative velocity, and the quiet surrounding Hubble flow represented through the positions and velocities of galaxies at distances from one to four~Mpc. Our approach delivers representative posterior samples of \LCDM realisations that are statistically and simultaneously consistent with all these observations, leading to significantly tighter mass constraints than found if the individual datasets are considered separately. In particular, we estimate the virial masses of the MW and M31 to be $\log_{10}(M_{200c}/M_\odot) = 12.07\pm0.08$ and $12.33\pm0.10$, respectively, their sum to be $\log_{10}(\Sigma M_{200c}/M_\odot)= 12.52\pm0.07$, and the enclosed mass within spheres of radius $R$ to be $\log_{10}(M(R)/M_\odot)= 12.71\pm0.06$ and $12.96\pm0.08$ for $R=1$~Mpc and $3$~Mpc, respectively. The M31-MW orbit is nearly radial for most of our \LCDM LG's, and most lie in a dark matter sheet that aligns approximately with the Supergalactic Plane, even though the surrounding density field was not used explicitly as a constraint. The approximate simulations employed in our inference are accurately reproduced by high-fidelity structure formation simulations, demonstrating the potential for future high-resolution, full-physics \LCDM posterior simulations of LG look-alikes. Such simulations would closely mirror the observed properties of the real system and its immediate environment.}


   \maketitle

\section{Introduction}
Abundant and very detailed observational information is available for the Local Group of galaxies. It thus provides a unique laboratory to study the physics of galaxy formation, galactic dynamics, and even cosmology. 
For instance, the Local Group has a complete sample of dwarf galaxies down to low luminosity, and the sample's properties have led to controversies like the ``Missing Satellite'' and ``Too-Big-to-Fail'' problems \citep[][and references therein]{bullockSmallScaleChallengesCDM2017} which highlight apparent differences between the observed dwarf luminosity function and that predicted by cosmological simulations.
This has driven refinements of galaxy formation modelling on small scales involving more realistic treatments of baryonic processes such as star formation and feedback, but it has also motivated the exploration of alternative particle physics candidates for dark matter \citep[][and references therein]{salesBaryonicSolutionsChallenges2022}.
Another observed peculiarity of the Local Group is the co-planar spatial and orbital configuration of the satellites surrounding the two major galaxies, which has led to debates on whether this can plausibly occur in a \LCDM universe \citep{libeskindDistributionSatelliteGalaxies2005,pawlowskiPlanesSatelliteGalaxies2018}. Some authors have argued 
that this configuration is transient and of low statistical significance \citep{sawalaMilkyWayPlane2023} and that the clustering of orbital planes may reflect coordinated infall of groups of satellites  \citep{liInfallSubstructuresMilky2008,taibiPortraitVastPolar2024}.
The Local Group is also the only place where we can use detailed kinematic and chemical information for individual stars to deduce the detailed assembly history of a galaxy, an area in which substantial progress has been made since the latest data release of \textit{Gaia} \citep{helmiStreamsSubstructuresEarly2020}.

Cosmological simulations are central to addressing these problems, as they provide realistic models for the formation of structure, which allows us to evaluate observations in the context of a specific physical model such as the \LCDM paradigm.
To understand whether particular observed properties of our Local Group are consistent with this paradigm, and to explore its predictions for the formation history of Local Group galaxies, it is valuable to construct a representative ensemble of \LCDM simulations tuned to match closely the well-observed properties of the two main galaxies and their environment.

Much effort has been made on modelling the formation of the Local Group, initially using two-body models \citep{kahnIntergalacticMatterGalaxy1959,gottAngularMomentumLocal1978,mishraPredictionsDistanceLocal1985}, later using Numerical Action Methods \citep{peeblesModelFormationLocal1989,peeblesGravitationalInstabilityPicture1990,peeblesVarietySolutionsDynamics2013}, and recently the community has developed several full cosmological simulation programmes aimed at the Local Group \citep[some recent ones include][]{gottloeberConstrainedLocalUniversE2010a,fattahiApostleProjectLocal2016,sawalaAPOSTLESimulationsSolutions2016,garrison-kimmelELVISExploringLocal2014,garrison-kimmelLocalGroupFIRE2019,libeskindHESTIAProjectSimulations2020,sawalaSIBELIUSProjectPluribus2022,mcalpineSIBELIUSDARKGalaxyCatalogue2022}. However, obtaining a representative sample of LG analogues that closely match relevant observed properties such as the masses of the two main galaxies, their separation and relative velocity, as well as their somewhat larger scale environment, has turned out to be challenging.
Many studies have started with a cosmological simulation with unconstrained initial conditions, and have picked out LG analogues in the simulation volume \citep{liMassesLocalGroup2008,gonzalezMASSLOCALGROUP2014,fattahiApostleProjectLocal2016,carlesiConstrainingMassLocal2017,zhaiLocalGroupAnalogs2020a,pillepichMilkyWayAndromeda2023}, but this approach has limited ability to match the Local Group at the present day because the number of analogues that one finds is directly related to the total parameter volume allowed by the selection criteria. Hence, 
tightening the criteria rapidly reduces the number of analogues found.
This can be partially mitigated by utilizing large volume simulations or by running many simulations with different random initial conditions to increase the volume surveyed, but finding a system that matches all desired properties of the Local Group accurately is not feasible. This can be seen in \cref{tab:previoussims} which provides the number of Local Group analogues found by different studies, as well as the strictness of the selection criteria. For example, we note that, in the APOSTLE suite, only 12 LG analogues were found to satisfy relatively loose criteria within a $(\SI{100}{Mpc})^3$ simulation volume.

\begin{table*}
    \centering
    \makebox[\linewidth][c]{
    \begin{tabular}{p{0.19\linewidth}p{0.12\linewidth}p{0.13\linewidth}p{0.13\linewidth}p{0.14\linewidth}p{0.15\linewidth}}
    \toprule
        & APOSTLE & ELVIS & Clues & Hestia & Sibelius ``Intermediate" (``Strict")\\
        \midrule
        Masses ($10^{12}\,\si{\Msun}$) & Total: 1.6 to 4 & Total: 2 to 5, each halo 1 to 3 & Total: \num{<7.1}, each halo \num{>0.71} & Each halo: 0.8 to 3 & Total: 1.5 to 5 (2 to 4)\\
        Mass ratio M31/MW  & - & - & - & 0.5 to 2 & $\frac{2}{3}$ to 3 (1 to 2) \\
        MW-M31 distance (\si{Mpc})  & 0.6 to 1.0 & 0.6 to 1.0 & 0.43 to 2.1 & 0.5 to 1.2 & 0.6 to 1.0 (0.74 to 0.8) \\
        Radial velocity (\si{km.s^{-1}}) & $-250$ to 0 & $<0$ (infalling) & - & $<0$ (infalling) & $-150$ to $-50$ ($-109$ to $-99$) \\
        Tangential vel. (\si{km.s^{-1}})  & \num{<100} & - & - & - & \num{<100} ( \num{<40})\\
        Isolation radius $r$ (\si{Mpc}) & 2.5 & 2.8 & 3.6 & 2 & 2 \\
        LSS constraints & - & - & Cosmicflows-II & Cosmicflows-II & \borg 2M++ \\
        Deviation w.r.t. LSS  & - & - & \SI{<7.1}{Mpc} & \SI{<3.5}{Mpc} & \SI{<5}{Mpc} \\
        Extra requirements & Hubble flow at \SI{2.5}{Mpc} within $\SI{\sim80}{km.s^{-1}}$ and slope within \SI{30}{km.s^{-1}.Mpc^{-1}} of LG values & No haloes with $M>\SI{7e13}{\Msun}$ within \SI{7}{Mpc} & Reduced angular momentum and energy within $2\sigma$ of observations. Embedded in a filament. & Virgo cluster with mass $\SI{>2e14}{\Msun}$, and no other more massive cluster within \SI{<20}{Mpc} of Virgo. & (Virgo present in selected \borg sample.) \\
        M31-MW axis deviation w.r.t. LSS & - & - & - & - & \SI{<30}{\degree} (\SI{<15}{\degree}) \\
        \midrule
        Number satisfying  & 12 in $(\SI{100}{Mpc})^3$ & 146 in 50 sims of $(\SI{70.4}{Mpc})^3$ & 42 / 300 (51 without filament criterion) & 13 / 1000 & 82 / 60000 (0 / 60000), without angle criterion 489 (1) \\
        Has hydro resimulations & yes & yes & no & yes & no \\
         \bottomrule
    \end{tabular}
    }
    \vspace{1em}
    \caption{A summary of the selection requirements of some notable past Local Group simulations from \citet{fattahiApostleProjectLocal2016,garrison-kimmelELVISExploringLocal2014,carlesiConstrainedLocalUniversE2016a,libeskindHESTIAProjectSimulations2020,sawalaSIBELIUSProjectPluribus2022}. For isolation, each simulation requires that there is no more massive halo than the smaller one of the pair within a sphere of radius $r$.}
    \label{tab:previoussims}
\end{table*}

A more direct and potentially more powerful approach is to use constrained simulations, in which one reconstructs initial conditions given some present-day constraints.
Two commonly applied methods exist. One is based on Wiener Filter reconstructions \citep{hoffmanConstrainedRealizationsGaussian1991,zaroubiWienerReconstructionLarge1999}, while the second and more recent is based on Bayesian inference, using Hamiltonian Monte Carlo methods.
Both have been very successful on large scales \citep{2002MNRAS.333..739M,courtoisThreedimensionalVelocityDensity2012,sorceCosmicflowsConstrainedLocal2016,tullyCosmicflows3CosmographyLocal2019,jaschePresentCosmicStructure2015a,jaschePhysicalBayesianModelling2019,lavauxSystematicfreeInferenceCosmic2019}, reproducing cosmic large-scale structures that match in detail those observed. Bayesian forward modelling schemes are generally preferable because the \citet{hoffmanConstrainedRealizationsGaussian1991} method only exactly samples from the posterior distribution for constraints that are linear in the initial condition field.
Bayesian forward modelling with HMC is more versatile in dealing with complex observational data and it can be extended to deal with non-linear scales.

So far, these techniques have only been applied with constraints based on datasets suited to constrain structure on a relatively large scale, such as galaxy redshift surveys, and peculiar velocity catalogues. They have never been applied to generate constrained simulations that match the properties of the Local Group and its immediate environment.  
For example, the 2019 analysis of 2M++ with \borg\citep{jaschePhysicalBayesianModelling2019} constrains realisations of cosmic structure to match the observed galaxy distribution binned in $(\SI{4}{Mpc})^3$ voxels, and constrained simulations based on peculiar velocity datasets \citep[e.g.][]{carlesiConstrainedLocalUniversE2016a} have a similar or larger effective filter scale, because of the requirement that the velocities are well-described by linear theory. Both methods are thus still far from what is required for a constrained simulation of the Local Group.

Several recent programmes use these large-scale constrained simulations as a base. They then do trial-and-error runs to identify Local Group analogues with roughly the observed properties.
This is done by first generating initial conditions using large-scale constraints. This sets the large-scale Fourier components of the initial density fields. The Fourier components on small scales remain unconstrained and are set at random. One then resimulates many different realisations and picks out those which produce a galaxy pair that is Local-Group-like.
For example, in the \emph{Clues} \citep{gottloeberConstrainedLocalUniversE2010a,carlesiConstrainedLocalUniversE2016a,sorceCosmicflowsConstrainedLocal2016} and more recent \emph{Hestia} simulations \citep{libeskindHESTIAProjectSimulations2020}, one starts from large-scale initial conditions based on a Wiener Filter reconstruction of peculiar velocity data from the CosmicFlows-2 catalogue \citep{tullyCOSMICFLOWS2DATA2013}.
Subsequently, for \emph{Hestia}, 1000 realisations with randomised small-scale initial conditions are created, resulting in 13 realisations with a halo pair that matches their relatively coarse Local Group criteria, as summarised in \cref{tab:previoussims}.

For the \emph{Sibelius} simulations, \citet{sawalaSIBELIUSProjectPluribus2022} go a step further, using for the large scale structure a realisation from the 2M++ galaxy count data-based reconstructions from \borg \citep{jaschePhysicalBayesianModelling2019}.
They also randomised the small-scale unconstrained structure and found a few good matches, although they did not satisfy their full ``wish list" of LG analogue selection criteria, which is briefly listed in \cref{tab:previoussims} and is more strict than that for the HESTIA simulations.
To get higher-fidelity realisations, they refined promising sets of initial conditions by iteratively randomising smaller and smaller scales, at each stage picking the ``best" simulation and then refining even smaller scales.
With some careful selection of promising candidates, this allowed them to obtain a small number of initial condition sets that match their strictest Local Group requirements (see \cref{tab:previoussims}).
However, with such a hierarchical, manual approach, the results cannot be assumed to be an unbiased and fair sample of the posterior distribution of \LCDM universes subject to the adopted Local Group constraints.

Such a sample of representative histories for the Local Group would however be extremely valuable.
One could then infer the expected distribution of Local Group properties in a \LCDM universe, conditional on its present-day configuration. 
For instance, one could check expectations for anisotropies in the satellite distribution (i.e. satellite planes), for the immediate environment of the Local Group, for the distribution of mass on various scales, for the inter-group and circum-group baryon distribution, and even for the detailed assembly history of the system.

In this work, we will use the Hamiltonian Monte Carlo (HMC) methods already used successfully on larger scales to generate constrained simulations of the Local Group \citep{jascheBayesianPhysicalReconstruction2013,lavauxUnmaskingMaskedUniverse2016,jaschePhysicalBayesianModelling2019}.
These methods come with some difficulties, however.
Notably the HMC algorithm requires the availability of an efficient evaluation for the derivatives of the posterior probability. In other words, one needs to be able to back-propagate the gradient of the likelihood to the cosmological initial conditions.
Additionally, the HMC requires generating many simulations. Keeping computational time feasible requires individual simulations to be efficient.
Because of this, constrained simulations cannot be built directly on top of state-of-the-art N-body simulation codes like \emph{Gadget-4} \citep{springelSimulatingCosmicStructure2021} or \emph{Swift} \citep{schallerSwiftModernHighlyparallel2023} without major modification.
Instead, the relevant community has developed custom cosmological simulation codes with hand-coded derivatives, usually based on Lagrangian perturbation theory \citep{jascheBayesianPhysicalReconstruction2013,kitauraInitialConditionsUniverse2013,wangReconstructingInitialDensity2013}, or using Particle Mesh (PM) gravity \citep{wangELUCIDExploringLocal2014,jaschePhysicalBayesianModelling2019,liPmwdDifferentiableCosmological2022}.
Particle mesh simulations are particularly attractive because they can be much improved and with marginal additional cost when utilising LPT-based knowledge in the integration, by the use of either \emph{COLA} \citep{tassevSolvingLargeScale2013}, as is done inside \borg, or \emph{FastPM} \citep{fengFASTPMNewScheme2016} as in e.g. \citet{modiFlowPMDistributedTensorFlow2021}. 
For our purposes, the existing integrator in \borg is not, however, sufficient to achieve the required resolution.

This paper proceeds with \cref{sec:methods}, where the HMC methodology for finding a representative sample of Local Group initial conditions is described, together with the observational data on the Local Group that we will use to constrain our inference. Then, in \cref{sec:results}, we present the results of this inference and the properties of the realistic and representative sample of Local Groups that it produces. In \cref{sec:resimulations} we explore how the initial conditions can be used to perform resimulations with Gadget-4. In \cref{sec:discussion} we discuss how our results compare to previous studies, we highlight some of its limitations, and we sketch some future research directions. Finally, in \cref{sec:summary} we conclude.

\section{Methods and setup}
\label{sec:methods}
\begin{figure*}
    \centering
    \includegraphics[width=\linewidth]{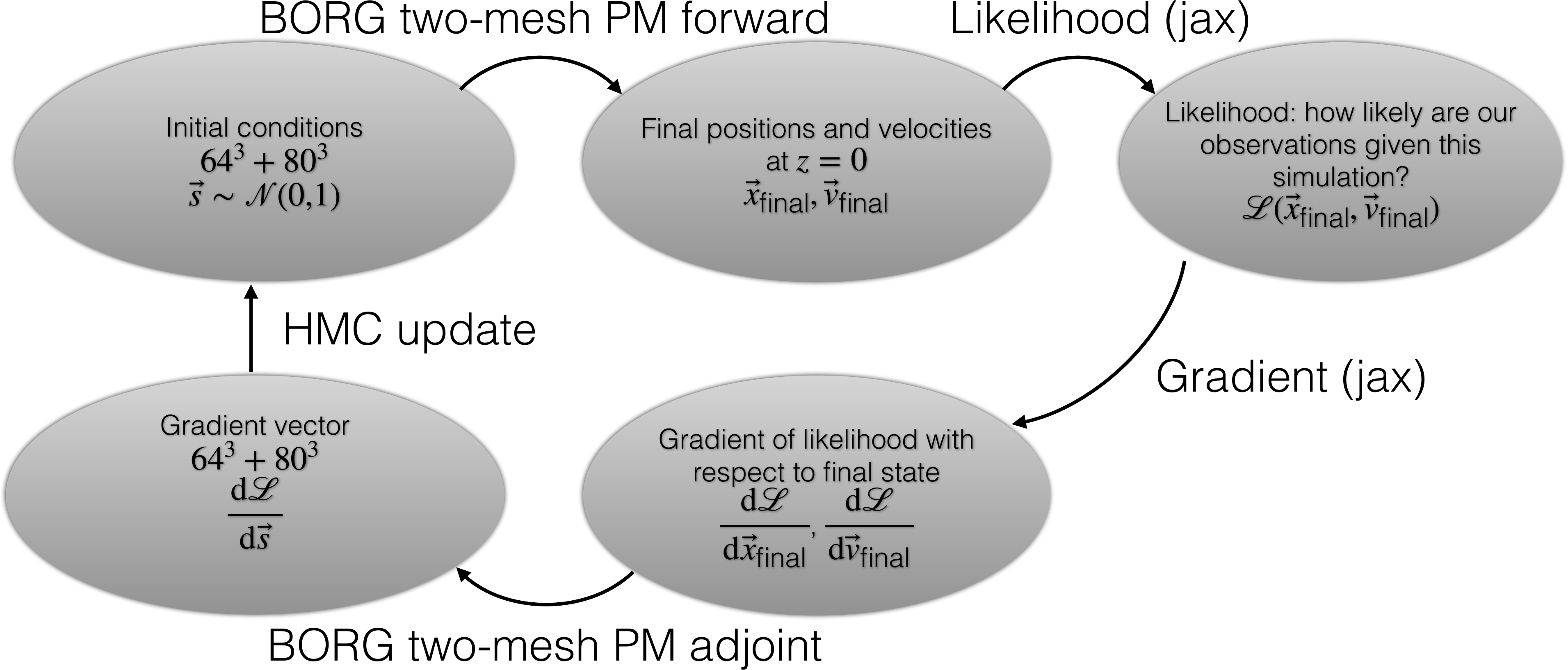}
    \caption{A schematic diagram of our Hamiltonian Monte Carlo method.}
    \label{fig:diagram}
\end{figure*}

For this work, we wish to obtain a fair statistical sample of \LCDM simulations, all of which reproduce with high fidelity the main observed characteristics of the two largest Local Group galaxy haloes, as well as the smooth and weakly perturbed Hubble flow in their immediate environment. 
Since a simulation is determined fully by forward integration from its initial conditions, obtaining a fair sample of simulations is equivalent to obtaining a fair sample of initial condition sets conditional on well-established observational constraints on the present-day system.
This motivates us to set up a Bayesian inference problem: we wish to sample the initial conditions, following a \LCDM prior, conditional on a likelihood that describes our observational knowledge of the Local Group and its environment.
The posterior initial conditions will then generate a fair statistical sample of simulations of LG analogues.
To build this machinery, we split this section into three parts: in \cref{sec:methodsinference}, we describe the inference process, which uses Hamiltonian Monte Carlo, in \cref{sec:forwardmodel}, we describe the forward model, which is the cosmological simulation from a Gaussian random initial density field to the present-day particle distribution, and in \cref{sec:lglik}, we describe the likelihood we adopt to summarise the observational knowledge we have of the Local Group and its immediate environment.

\subsection{Bayesian inference of Large Scale Structure formation}
\label{sec:methodsinference}
In cosmological simulations, the initial conditions on particles are created by Gaussian random numbers $\vb*{s}$, which, assuming a power spectrum and Lagrangian perturbation theory, are then used to determine the initial high-redshift particle positions and velocities \citep[][and references therein]{anguloLargescaleDarkMatter2022}.
We wish to carry out Bayesian inference on the Gaussian white noise field that generates the initial conditions of our simulation, sampling from the posterior 
\begin{equation}
  \log{P}(\vb*{s}) = \log{\pi}(\vb*{s}) + \log{\mathcal{L}(F(\vb*{s}))},
\end{equation}
where $\pi$ is the prior probability,  $\vb*{s}$ indicates the Gaussian random field that generates the initial conditions, $F(\vb*s)$ is the result of the forward model, i.e. our particle-mesh simulations, and $\mathcal{L}$ indicates the likelihood function which describes how likely the observed properties of the Local Group are, given the result of a simulation (see \cref{sec:lglik}). 

\subsubsection{Hamiltonian Monte Carlo}
For the inference, we utilise Hamiltonian Monte Carlo \citep[HMC, ][]{duaneHybridMonteCarlo1987,nealProbabilisticInferenceUsing1993}, a sampling method that works well for high-dimensional problems and has been successfully used to make large-scale structure inference \citep{jascheBayesianPhysicalReconstruction2013,lavauxUnmaskingMaskedUniverse2016,jaschePhysicalBayesianModelling2019}. 
The process is well-described in \citet{brooksHandbookMarkovChain2011}, and we summarise it briefly here. Conceptually, it means solving numerically the equations of motion for a fictitious physical system, where the potential is minus the log-posterior of the problem. 
In our case, we consider the motion of our initial condition field vector $\vb*{s}$, inside the potential defined by the log-posterior of our problem:
\begin{align}
    H(\vb*{s}, \vb*p) &= -\log{P(\vb*s}) + \frac{1}{2} \vb*p^T \vb*M^{-1} \vb*p\\
    \dv{\vb{p}}{\tau} &= -\pdv{H}{\vb*s} = \grad\log{P}(\vb*s)\label{eq:eomp}\\
    \dv{\vb{s}}{\tau} &= \pdv{H}{\vb*p} = \vb*M^{-1} \vb*p \label{eq:eoms}
\end{align}
where $\vb*p$ is the momentum and $\vb*M$ is the mass matrix. Each iteration of the Hamiltonian Monte Carlo sampler consists of three steps. One first (step 1) integrates the equations of motion (\cref{eq:eomp,eq:eoms}) for a given number of steps $N_\text{steps}$ using a step size $\epsilon$. The integrator must be symplectic.
Then (step 2) after the integration, the proposed result is rejected or accepted based on a Metropolis-Hastings criterion: 
\begin{align}
    p_\text{accept} &= \min(1, \exp(-\Delta H))\\
    \Delta H &= H(\vb*s_\text{proposed}, \vb*p_\text{proposed}) - H(\vb*s_\text{original}, \vb*p_\text{original}),
\end{align}
and the new state becomes $(\vb*s_\text{proposed}, \vb*p_\text{proposed})$ upon acceptance, and $(\vb*s_\text{original}, -\vb*p_\text{original})$ upon rejection.
And lastly (step 3), the momentum is refreshed
\begin{equation}
    \vb*p \to \alpha \vb*p + \sqrt{1-\alpha^2} \mathcal{N}(\vb*\mu=\vb*0,\vb*\Sigma=\vb*M),
\end{equation}
where $\mathcal{N}(\vb*\mu=\vb*0,\vb*\Sigma=\vb*M)$ is a multivariate Gaussian distribution with mean $\mu$ and covariance matrix $\vb*\Sigma$.

This gives the HMC algorithm several hyperparameters that must be chosen. 
We opt for a leapfrog scheme to integrate Hamilton's equations of motion, with a fixed number of steps $N_\text{steps}$, and where the step size drawn is uniformly between $[0.9\epsilon_\text{max},\epsilon_\text{max}]$.
We set the mass matrix $\vb*M$ to be an identity matrix, and we set the momentum conservation ratio $\alpha$ to zero; that is, we do a full momentum refreshment each step. 

The parameters $\epsilon_\text{max}$ and $N_\text{steps}$ have to be tuned, and we tune them to achieve acceptance rates around \SIrange{50}{70}{\percent}, while maximizing the integration length per step -- this is required to keep random-walk behaviour to a minimum \citep{brooksHandbookMarkovChain2011}.
Although adopting a different mass matrix can help sampling performance, it is not practical to find one, especially in high-dimensional problems like ours.
Similarly, tuning $\alpha$ can also benefit sampling performance in some situations; we found no measurable improvement in some exploratory tests, so we kept it to zero.
Lastly, we have tried replacing the leapfrog integrator with higher-order schemes as suggested by \citet{hernandez-sanchezHigherOrderHamiltonian2021}, but some cursory tests showed a simple leapfrog integration to perform the best for this problem. More complete testing is left for future work. 

\subsubsection{Two-grid sampling}
In our procedure, we use a zoom setup, in which our Gaussian white noise field $\vb*{s}$ is actually composed of two white noise fields for a low-resolution (LR) grid $\vb*{s}_\text{LR}$ and a high-resolution (HR) grid $\vb*{s}_\text{HR}$. These are $N_\text{LR}^3$ and $N_\text{HR}^3$ grids with each element being a $\mathcal{N}(0,1)$ standard Gaussian random variable. In our procedure, we jointly sample these fields.

\subsection{The forward model}
\label{sec:forwardmodel}
In our case, the forward model is the gravity simulation that takes in the Gaussian white noise fields $\vb*{s}$ and outputs the particle positions and velocities at the present time. 
Four steps are involved. (i) We convolve the Gaussian white noise field $\vb*{s}$ with (the square root of) the (appropriately scaled) power spectrum, in our case the linear matter power spectrum $P(k)$ with the backscaling method \citep{anguloLargescaleDarkMatter2022}. This gets us the initial condition of the matter field at $z=63$. (ii) We compute the Lagrangian displacement field given the generated Gaussian random field at $z=63$ and interpolate this displacement onto our particle load that is initially placed on two cartesian grids. In particular, we place high-resolution particles at the centres of each voxel of the high-resolution grid and place low-resolution particles at the centres of each voxel in the low-resolution grid that is not in the high-resolution region. This results in $64^3$ high-resolution particles and $64^3-16^3$ low-resolution particles\footnote{Since the zoom region has a side length of $1/4$ of the full box, $(64/4)^3 = 16^3$ low-resolution particles are replaced by the high-resolution particles.}. (iii) We perturb the positions and velocities using the Lagrangian displacement field to obtain the particle initial conditions at $z=63$. (iv) We run a gravity simulation given these particles down to $z=0$, for which we use a two-level particle-mesh method.

\subsubsection{Particle mesh simulations}
To compute particle forces, in the usual single-mesh particle mesh simulations \citep{hockneyComputerSimulationUsing1988}, one also has to do four steps. (i) We compute the density field by depositing particles onto a mesh, distributing the mass of each particle on the nearest mesh cells with the piecewise linear Cloud-In-Cell (CIC) kernel. (ii) We obtain the potential field from this density field by solving the Poisson equation on a mesh. In practice, we convolve the density field with the Green's function in Fourier space (this is denoted as $\Delta^{-1}$ because, on the mesh, this operation can be thought of as the inverse of the Laplacian). (iii) We compute a force field from the potential field by finite differentiation.  (iv) We interpolate the forces onto the particles using a CIC kernel.
When taking derivatives of the potential, which must be consistent with the choice of Green's function, we use a 2nd-order finite-difference approximation of the gradient compatible with our treatment of the inverse Laplacian \citep[see][for an overview of alternative higher-order approximations]{hahnMultiscaleInitialConditions2011}.
Using the forces, the equations of motion are then integrated in time using a leapfrog integrator. 
In this work, we build on the implementation existing in \borg, as described in \citet{jaschePhysicalBayesianModelling2019}.

The computational effort of particle deposition onto a density grid scales with the number of particles, and the computational effort of the Fourier transform with the number of voxels $\mathcal{O}(N\log{N})$, where $N=(L/\Delta x)^3$, with $L$ the box size and $\Delta x$ the voxel size. Thus, the computational effort is determined by the required box size, resolution and particle mass. 

\subsubsection{Zoom simulations}
\label{sec:methodszoomsimulations}
In practice, single-mesh simulations result in a rather limited dynamic range. This is why many improved algorithms have been proposed, for instance, tree-based codes, codes using the fast multipole method, and zoom methods. Here we will use a zoom method because we require high enough spatial resolution to resolve the haloes of the MW and M31 (which have virial radii of order \SI{200}{kpc}) within a region large enough to avoid artefacts from imposed periodic boundary conditions. 
In our zoom simulations, a fixed-position high-resolution (HR) zoom region populated with high-resolution (i.e. low-mass) particles is embedded within a lower resolution (LR)  but larger region populated with more massive particles.

The particular zoom method we employ is a relatively simple extension of the single-grid particle-mesh method. 
The forces are split into two parts: a long-range and a short-range part, where the long-range force is computed from the LR grid, and the short-range force is computed from the HR grid.
The force on each particle $i$ is
\begin{align*}
    \vb{F}_i = 
    \begin{cases}
    \vb{F}_\mathrm{HR}(\vb{x}_i) + \vb{F}_\mathrm{LR}(\vb{x}_i) &\qq{if $\vb{x}_i$ is inside the HR region}\\
    \vb{F}_\mathrm{LR}(\vb{x}_i) &\qq{if $\vb{x}_i$ is outside the HR region,}
    \end{cases}
\end{align*}
where $\vb{F}_\mathrm{LR}(\vb{x})$ is the tri-linearly interpolated force vector at position $\vb{x}$ of the LR region's force field, and $\vb{F}_\mathrm{HR}(\vb{x})$ is the equivalent interpolated high-resolution force within the HR region.

To carry out the force splitting faithfully, we convolve the LR potential with a Gaussian filter,
\begin{equation}
    K_\text{smooth}(\abs{\vb*k}) = \exp(-\abs{\vb*k}^2 (A_\text{smth} L/N)^2),
\end{equation}
where we set $A_\text{smth} = 1.25$, which is the default value when matching the PM and tree forces in Gadget-4 \citep{springelSimulatingCosmicStructure2021}. We also convolve the HR potential field with a high-pass filter which is the complement of this Gaussian filter, such that at each frequency in $k$-space, the total power is preserved.
Additionally, since some power is lost in the low-resolution force field as a result of CIC deposition and interpolation, we recover this power from the HR field. 
We also zero-pad the HR grid in order to achieve vacuum boundary conditions. We found that we require 16 cells of padding, corresponding to eight cells on each side, in order to have convergence on the final positions of the MW-M31 pair.

The final result is that our `effective' gravitational potential -- the potential from which the force fields  $\vb{F}_\text{LR}$ and $\vb{F}_\text{HR}$ are calculated -- is, in Fourier space
\begin{multline}
    \widetilde{\vb*{\Phi}}_\mathrm{LR}(\vb*k) = \tilde\delta_\text{LR} \times \Delta^{-1}_\text{LR}(\vb*k) \times K_\text{smooth}(\abs{\vb*{k}}) \\ 
    \widetilde{\vb*{\Phi}}_\mathrm{HR}(\vb*k) = \tilde\delta_\text{HR} \times (\Delta^{-1}_\text{HR}(\vb*k) \\- \Delta^{-1}_\text{LR}(\vb*k) \times K_\text{CIC,LR}^2(\vb*k) \times K_\text{smooth}(\abs{\vb*{k}})),
\end{multline}
where $\delta_\text{LR}$ is the density field of the low-resolution region, deposited with the grid size of the low-resolution region, and $\Delta^{-1}$ is the inverse Laplacian.
It is understood that $K_\text{CIC,LR}$ is set to be zero beyond the Nyquist frequency of the LR grid. 

For the initial conditions, we consider two initial Gaussian random fields, $\vb*\Phi_\text{LR}(k)$ and a $\vb*\Phi_\text{HR}(k)$. These are both simply white noise fields convolved with the matter power spectrum with the appropriate linear scaling. From these we obtain the Lagrangian displacement fields $\vb*\Psi_\text{LR, i}$ and $\vb*\Psi_\text{HR, i}$, similarly as the force fields, except that now the total of the squares of the two filters must sum to one \citep[equation 5 in][]{stopyraGenetICNewInitial2021}. This results in the displacement fields 
\begin{multline}
    \tilde{\vb*{\Psi}}_\mathrm{LR, i}(k_x, k_y, k_z) = \phi_\text{LR} \times \frac{k_i}{\abs{\vb*{k}}^2} \times K_\text{smooth}(\abs{\vb*k})\\ 
    \tilde{\vb*{\Psi}}_\mathrm{HR, i}(k_x, k_y, k_z) = \phi_\text{HR} \times \frac{k_i}{\abs{\vb*{k}}^2} \times \\\sqrt{1 - (K_\text{CIC,LR}(k_x, k_y, k_z)\times K_\text{smooth}(\abs{\vb*{k}}))^2}.
     \label{eq:zoomlpt}
\end{multline}
To get the actual particle displacements, we do a CIC interpolation of the initial particle grids onto this grid.

Because we use this zoom simulation algorithm inside a Hamiltonian Monte Carlo sampler, we also need to derive its adjoint gradient. This calculation is analogous to Appendix~C of \citet{jaschePhysicalBayesianModelling2019}, with the inclusion of the different particle masses for the low- and high-resolution particles; this is just an extra constant factor on the mass deposition end. 
Additionally, the force adjoint gradient will now have contributions from both the LR and HR grids. 

The parallelisation strategy is also similar to \borg.
In the single-mesh particle-mesh code of the \borg framework, each MPI task handles a slab of the full simulation box as well as the particles in this slab.
At each timestep, we transfer particles that move into a different slab to their new MPI task. 
This allows for simple parallelisation of the FFTs as well as the CIC assignment and interpolation. 
In our zoom algorithm, the process is similar, except that each MPI node owns both a slab of the LR box and a slab of the HR box.

\subsection{The Local Group likelihood}
\label{sec:lglik}
The likelihood indicates how ``Local-Group-like" the final state of a simulation is, or, more precisely, it quantifies the level of consistency between the observational constraints and each particular simulated Local Group analogue, given the observational uncertainties.  
To define this likelihood, we first introduce the concept of filtered masses, positions and velocities in \cref{sec:filtquantities} as a simple means to extract robust and differentiable values for such quantities from the particle data of our simulation. 
Our total Local Group likelihood is 
\begin{equation}
\mathcal{L}_\text{LG} = \mathcal{L}_\text{mass} + \mathcal{L}_\text{position} + \mathcal{L}_\text{velocity} + \mathcal{L}_\text{flow}.
\end{equation} 
We include observational constraints on the masses and position of the two haloes ($\mathcal{L}_\text{mass}$ and $\mathcal{L}_\text{position}$, \cref{sec:likmass}). We also constrain the relative MW-M31 velocity vector  ($\mathcal{L}_\text{velocity}$, \cref{sec:likvelocity}). Finally, we require that the local Hubble flow as traced by individual galaxies in the Local Group's immediate surroundings should be reproduced within the observational uncertainties ($\mathcal{L}_\text{flow}$, \cref{sec:localflow}).

\subsubsection{Filtered masses, positions and velocities}
\label{sec:filtquantities}
To be able to compare the output of our simulation to observations, we need to consider some quantities that are easy to measure both in simulations and through observation. 
We consider therefore the masses, positions and velocities of the objects of interest, averaged over some fixed filter. 
The filtered mass is defined as a function of position,
\begin{equation}
    M(\vb{x}_\text{filter}, \sigma) = \int_{\mathbb{R}^3}{d^3\vb{r} \, \rho(\vb{r})} G(\vb{r} - \vb{x}_\text{filter}, \sigma),\label{eq:filtmass}
\end{equation}
where $G(\Delta\vb{r},\sigma) = e^{-\Delta\vb{r}^2/2\sigma^2}$ is an isotropic Gaussian filter with standard deviation $\sigma$.
Although it might seem desirable to use a tophat filter so that one can use observationally inferred enclosed masses, the result would not be differentiable with respect to the particle positions, which would make the gradients non-informative and reduce the HMC's efficiency.
In the case of particles at locations $\vb{r}_i$, $\rho(\vb{r}) \approx\sum_i m_i\delta(\vb{r}-\vb{r}_i)$, so this becomes:
$$M(\vb{x}_\text{filter}, \sigma) = \sum_i m_i G(\vb{r}_i-\vb{x}_\text{filter},\sigma),$$
where $m_i$ is the mass of particle $i$.
We also constrain the offset of the filtered centre of mass from the filter centre.
\begin{align} \Delta\vb{x}_\text{CoM}(\vb{x}_\text{filter}) &= \frac{\int{d\vb{x}\ (\vb{x} - \vb{x}_\text{filter}) \rho(\vb{x}) G(\vb{x}-\vb{x}_\text{filter})}}{\int{d\vb{x}\rho(\vb{x}) G(\vb{x}-\vb{x}_\text{filter})}}\\
&=\frac{\sum_i{m_i (\vb{x}_i - \vb{x}_\text{filter}) G(\vb{x}_i-\vb{x}_\text{filter})}}{\sum_i{m_i G(\vb{x}_i-\vb{x}_\text{filter})}}. \label{eq:comfilt}
\end{align}
Lastly, we can define the filtered velocity, which is the average velocity within the filter:
\begin{equation}
\vb{v}_\text{filt}(\vb{x}_\text{filter}) = \frac{\sum_i{m_i \vb{v}_i G(\vb{x}_i-\vb{x}_\text{filter})}}{\sum_i{m_i G(\vb{x}_i-\vb{x}_\text{filter})}}.\label{eq:velfilt}
\end{equation}

\subsubsection{Mass and position}
\label{sec:likmass}
Observational estimates of the masses of the Milky Way and Andromeda give rise to the mass likelihood $\mathcal{L}_\text{mass}$, which consists of the terms\footnote{In this work, we use the notation $x \sim \mathcal{N}(\mu, \sigma^2)$ to mean that $x$ is distributed like a normal distribution with mean $\mu$ and standard deviation $\sigma$. In practice, quantities of this form appear as quadratic terms in the log-likelihood of our inference problem.} \begin{align}
    \ln(M(\vb{x}_\text{MW,obs}, \SI{100}{kpc})) &\sim \mathcal{N}(\ln(M_\text{MW,obs,100}), \sigma^2_\text{MW,obs,100,rel}) \label{eq:likMMW}\\
    \ln(M(\vb{x}_\text{M31,obs}, \SI{100}{kpc})) &\sim \mathcal{N}(\ln(M_\text{M31,obs,100}), \sigma^2_\text{M31,obs,100,rel}) \label{eq:likMM31},
\end{align}
where $M_\text{MW,obs}$ is the observational estimate of the \SI{100}{kpc} filtered mass of the Milky Way and $M_\text{M31,obs}$ that of M31. $\sigma_\text{MW,obs,rel}$ and $\sigma_\text{MW,obs,rel}$ are the associated relative errors.
The filter size of \SI{100}{kpc} was chosen because halo masses are well converged on this scale at our simulation resolution and because observational estimates on this scale have reasonably low uncertainty. 
The estimates we use come from the joint fitting of several observational tracers of the mass profiles of the Milky Way and Andromeda.
The data used and assumptions made to obtain them are described in \cref{app:observationalmasses}, and the final values with associated uncertainties are summarised in \cref{tab:massestimatesadopted}. We note that the uncertainties are mainly the observational uncertainties, but we have also added another
 \SI{5}{\percent} in quadrature to account for halo-to-halo scatter in the relation used to convert the mass found for our fitted contracted NFW profiles to the equivalent mass for a dark-matter-only halo \citep[we follow the recipe of ][]{cautunMilkyWayTotal2020}.

\begin{table*}
  \begin{center}
    \begin{tabular}[c]{llp{8cm}}
      \toprule
      Constraint & Functional form & Values \\
      \midrule
      Milky Way filtered mass & $ \ln(M(\vb*x_\text{obs}, w)) \sim \mathcal{N}(\ln(M_\text{obs}), \sigma_\text{rel}^2) $ & $\vb*x_\text{obs}=(-3.9, -6.9, -1.8)~\si{kpc}$, w = $\SI{100}{kpc}$, $M_\text{obs} = \SI{1.085e12}{\Msun}$, $\sigma_\text{rel} = \sqrt{0.05^2+0.1291^2}$ \\
      Milky Way centre-of-mass & $\Delta\vb{x}_\text{CoM}(\vb{x}_\text{obs})\sim\mathcal{N}(\vb*0, \vb*\Sigma=\vb*\Sigma_\text{x,res})$ & $\vb*\Sigma_\text{x,res} = (\SI{30}{kpc})^2 \Identity$\\
      \midrule
      M31 filtered mass & $ \ln(M(\vb*x_\text{obs}, w)) \sim \mathcal{N}(\ln(M_\text{obs}), \sigma_\text{rel}^2) $ & $\vb*x_\text{obs}=(780,0,0)~\si{kpc}$, $w = \SI{100}{kpc}$, $M_\text{obs} = \SI{1.288e12}{\Msun}$, $\sigma_\text{rel} = \sqrt{0.05^2+0.167^2}$ \\
      M31 centre-of-mass & $\Delta\vb{x}_\text{CoM}(\vb{x}_\text{obs})\sim\mathcal{N}(\vb*0, \vb*\Sigma=\vb*\Sigma_\text{x,res})$ & $\vb*\Sigma_\text{x,res} = (\SI{30}{kpc})^2 \Identity$\\
      \midrule 
      M31-MW velocity & \cref{eq:vdiffm31mw} & $\Delta \vb*v/(\si{km.s^{-1}}) = (-114.7,   39.8,  -59.8), \vb*\Sigma_\text{meas}/(\si{km.s^{-1}})^2 =  (0.9^2,39.2^2,30.2^2)\Identity, \vb*\Sigma_\text{v,res}=(\SI{10}{km.s^{-1}})^2\Identity$\\
      \midrule
      Local Hubble flow & \cref{eq:vhelio} & 31 galaxies, $\sigma_\text{int} = \SI{35}{km.s^{-1}}$, $w=\SI{250}{kpc}\cdot (d/\si{Mpc})$ \\
      \bottomrule
    \end{tabular}
  \end{center}
  \caption{A summary of our adopted observational constraints and of the parameter values that are used in building the likelihood, as described in \cref{sec:lglik}. The final likelihood also assumes an isotropic Gaussian uncertainty with dispersion \SI{10}{km.s^{-1}} per component in the reference frame corresponding to the velocity of the MW halo (see \cref{sec:veloffsetframe}).}
  \label{tab:massestimatesadopted}
\end{table*}

We use a heliocentric coordinate frame that is rotated to have M31 on the $x$-axis, that is $
\vb{x}_\text{M31,obs} = (780,0,0)\,\si{.kpc}$ (our assumed distance to M31 is \SI{780}{kpc}, see \cref{app:coordinates}).
In this coordinate frame, $\vb{x}_\text{MW,obs}\approx (-3.9, -6.9, -1.8)\,\si{kpc}$.

Our mass likelihood (\cref{eq:likMMW,eq:likMM31}) is insufficient on its own, because the observed filtered mass could be matched by an overly massive but substantially offset halo. 
We therefore introduce an additional likelihood term, $\mathcal{L}_\text{position}$, that forces the filtered centre of mass to be close to the filter centre. Specifically, we constrain the offset in position to be consistent with
\begin{align}
    \Delta\vb{x}_\text{CoM}(\vb{x}_\text{MW, obs})&\sim\mathcal{N}(\vb*0, \vb*\Sigma=\vb*\Sigma_\text{x,res})\label{eq:likcommw}\\
    \Delta\vb{x}_\text{CoM}(\vb{x}_\text{M31, obs})&\sim\mathcal{N}(\vb*0, \vb*\Sigma=\vb*\Sigma_\text{x,res}),\label{eq:likcomm31}
\end{align}
where we set the covariance matrix $\vb*\Sigma_\text{x,res}$ to be isotropic with dispersion $\sigma$ of \SI{30}{kpc} per direction, which allows for some spread in the simulations but does not lead to significant bias in estimations of the mass.

\subsubsection{Velocity likelihood}
\label{sec:likvelocity}
We also wish to use the relative velocity of the Milky Way and M31 as a constraint.
In our setup, we assume the distance between the Milky Way and M31 to be known. Therefore we translate the measured proper motion vector with its uncertainties into a transverse velocity vector with associated uncertainties. 
This allows us to describe the likelihood $\mathcal{L}_\text{velocity}$ for the 3D relative velocity of the barycentres of the two haloes as a 3D multivariate Gaussian:
\begin{align}
\vb{v}_\text{filt}(\vb{x}_\text{M31}) - \vb{v}_\text{filt}(\vb{x}_\text{MW}) \sim \mathcal{N}(\vb{v}_\text{meas}, \vb*\Sigma=\vb*\Sigma_\text{meas} + \vb*\Sigma_\text{v,res}), \label{eq:vdiffm31mw}
\end{align}
where $\vb*\Sigma_\text{meas}$ is the covariance matrix that incorporates the observational errors on the radial velocity and proper motions, and $\vb*\Sigma_\text{v,res}$ is some additional uncertainty that we include because adopting overly precise values for the constraints would result in inefficient sampling. Moreover, in reality, M31's 100-kpc filtered halo velocity may be offset from the galaxy's velocity.
We choose $\vb*\Sigma_\text{v,res}$ to be isotropic:  $\vb*\Sigma_\text{v,res} = (\SI{10}{km.s^{-1}})^2 \Identity$, where $\Identity$ is the identity matrix. 
We give the observational values adopted in \cref{app:coordinates}.

Although it would, in principle, be better to compare the proper motions directly to the simulated data, this does not matter in practice: the positions of the two haloes are allowed a dispersion of $\SI{30}{kpc}$, i.e. a $\SI{\sim 5}{\percent}$ scatter in the distance. The uncertainties on the transverse velocities due to proper motion uncertainties are, however, much larger.

\subsubsection{Local galaxy flow likelihood}
\label{sec:localflow}
Lastly, we utilise peculiar velocity measurements of galaxies in the immediate environment of the Local Group. 
Such measurements have shown that the Hubble flow is particularly quiet in our surroundings \citep{sandageStepsHubbleConstant1975,schlegelHowUnusualLocally1994,karachentsevHubbleFlowLocal2009,aragon-calvoUnusualMilkyWaylocal2023}. This has implications for the mass of the Local Group, as it shows that there is no large unresolved mass apart from the Milky Way and M31 in our neighbourhood and puts bounds on the possible total mass of the Local Group \citep{penarrubiaDynamicalModelLocal2014}.
This makes it an ideal constraint for our purposes because it naturally enforces a dynamically quiet environment around the MW and M31 without need for an explicit isolation criterion of the kind usually imposed when identifying LG analogues in simulations \citep{gottloeberConstrainedLocalUniversE2010a,libeskindHESTIAProjectSimulations2020,sawalaSIBELIUSProjectPluribus2022}.
In particular, we constrain the radial component of the velocity at the location of our sample of galaxies (which we describe below). Each galaxy gives us the following constraint for the flow likelihood $\mathcal{L}_\text{flow}$:
\begin{align}
    v_\text{helio} &\sim \mathcal{N}(v_\text{meas}, \sigma^2_v)\label{eq:vhelio}\\
    \sigma_v &= \sqrt{(H_0\sigma_\text{d})^2+\sigma_\text{int}^2}\\
    \sigma_d&=\frac{\ln{10}}{5} d\sigma_{\mu}
\end{align}
where $v_\text{helio}$ is the simulation-derived velocity at the galaxy's location converted to the heliocentric frame, $v_\text{meas}$ is the measured heliocentric velocity, $d$ is the distance, $\sigma_d$ is the (linearised) uncertainty on the distance, $\sigma_\mu$ is the error on the distance modulus, and $\sigma_\text{int}$ is some intrinsic scatter reflecting residual peculiar velocities. To compute the velocity in heliocentric frame $v_\text{helio}$, we evaluate the 3D filtered velocities at the locations of the galaxy in question, as well as the 3D filtered velocity at the location of our Milky Way (from \cref{eq:velfilt}). 
We compute from their difference a simulated Galactocentric velocity and then convert this to a heliocentric frame.

In order to compute filtered velocities for distant galaxies, we use a filter size of $(\SI{250}{kpc}) \times (d/\si{Mpc})$, where $d$ is the distance. The increasing filter size ensures that we have enough particles inside our filter, also at larger distances where we might only have a few particles because the corresponding Lagrangian region can be (partly) outside the zoom region.

We require the residual intrinsic scatter $\sigma_\text{int}$ in our setup because we only put in the Milky Way and M31 as matter density constraints.
Thus, many structures that contribute to the peculiar velocities of more distant galaxies are missing in our simulations.
Some of these structures could, in principle, be included, at the cost of making our inference more complex.
However, our simulations do not resolve lower mass structures and the observational data on known higher mass structures such as the M81, Centaurus and Maffei groups are uncertain; hence we prefer, for now, to ignore them.
We adopt $\sigma_\text{int}=\SI{35}{km.s^{-1}}$ since \citet{penarrubiaDynamicalModelLocal2014} estimated $\sigma_\text{int}=\SI[parse-numbers  = false]{35^{+5}_{-4}}{km.s^{-1}}$ for the residual scatter around their spherical infall model, based on a very similar sample of galaxies.

To build our sample, we join the galaxies from two sources: the catalogue of local volume galaxies \citep{karachentsevUPDATEDNEARBYGALAXY2013},\footnote{\url{http://www.sao.ru/lv/lvgdb/}} and the CosmicFlows-1 catalogue \citep{tullyExtragalacticDistanceDatabase2009},\footnote{\url{https://edd.ifa.hawaii.edu/}} in order to get a complete and up-to-date sample which combines as many independent distance measurements per galaxy as possible. For measurement errors, we follow CosmicFlows-1, which assumes a distance modulus error of 0.2 for a TRGB-only measurement, 0.16 for a TRGB + Cepheid measurement, and similar for other distance measurement methods. In our sample, we include only isolated galaxies, because these are expected to trace well the underlying flow field. Galaxies that have an interaction with a nearby massive perturber would need to be modelled more carefully.

\begin{figure}[tpb]
    \centering
    \includegraphics[width=\linewidth]{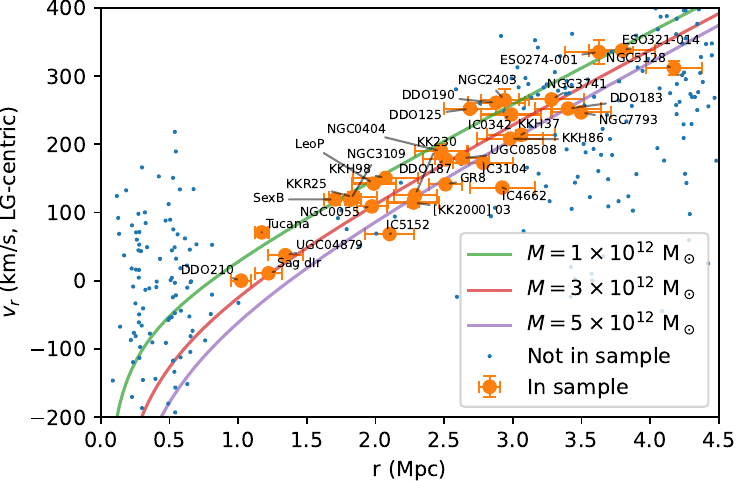}
    \caption{The catalogue of galaxies used as flow tracers in this analysis, analogous to Fig. 11 in \citet{penarrubiaDynamicalModelLocal2014}. The x-axis is the Local-Group-centric distance, where the centre of the Local Group is the centre-of-mass of the Milky Way-M31 pair when assuming a mass ratio of $M_\text{M31}/M_\text{MW}=2$. The y-axis is the radial velocity projected in the LG-centric frame, where the offset is computed by interpolating the velocity field along the line between the two galaxies linearly. The different lines indicate spherical Kepler-like velocity models of the recession velocity \citep[equation (9) of][]{penarrubiaDynamicalModelLocal2014}.}
    \label{fig:eddall}
\end{figure}

\begin{figure*}
    \centering
    \includegraphics[width=\linewidth]{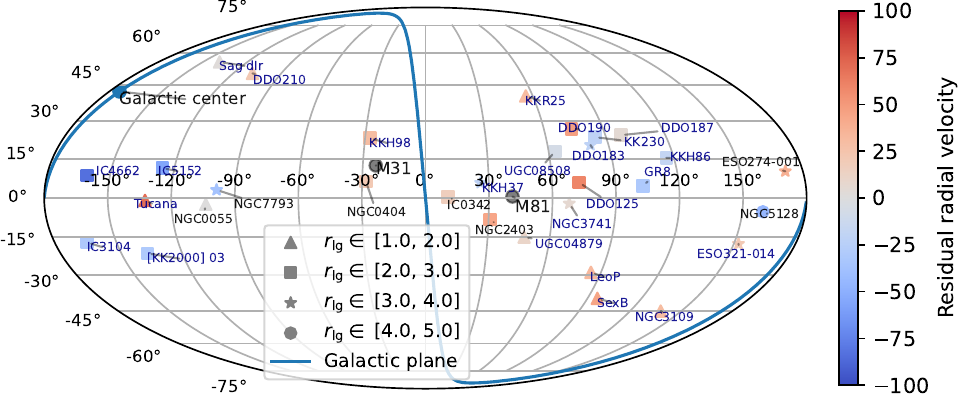}
    \caption{The sky distribution in Supergalactic coordinates of the catalogue of galaxies used as flow tracers in this analysis. Symbol type indicates distance from the Local Group's barycentre, and colour indicates velocity with respect to the spherical model shown as a red line in \cref{fig:eddall}. Units are in \si{Mpc} and \si{km.s^{-1}}. The blue line shows the Galactic plane. }
    \label{fig:eddsky}
\end{figure*}

Additionally, we only include galaxies with relatively small distance errors to ensure a reliable estimate of the velocity at the galaxy's position.
To be specific, we use the following criteria to build our sample of galaxies:
\begin{itemize}
\tightlist
\item
  The distance error must be less than \SI{0.25}{Mpc}. Given the error estimates we have adopted, this limits the sample to be within \SI{\sim4}{Mpc}. 
\item
  There must not be a more luminous galaxy within 0.5 Mpc that has $K < -15$. The 2MASS $K$ magnitude is used for this. If no K-band magnitude is measured, the estimate from \citet{karachentsevUPDATEDNEARBYGALAXY2013} is used.
\item
  We exclude the data points for the MW and M31 because they are already treated.
\item
  We exclude the data point for M81, because of its interaction with M82 and NGC 3077. The three galaxies are
  strongly interacting and their velocities cover a range of more than \SI{200}{km.s^{-1}}, so including them would require a more thorough dynamical analysis of the group.
\item
  To filter out satellites, we exclude any galaxy within \SI{800}{kpc} of: The Milky Way, Andromeda, M81, Maffei 1 / IC 342, Centaurus A (NGC5128).
\item 
  The measurement error on the velocity must be less than \SI{20}{km.s^{-1}}. In practice, this cuts out one galaxy (ESO006-001) which has a radial velocity uncertainty of \SI{58}{km.s^{-1}}.
\end{itemize}
We are left with a sample of 31 galaxies. 
Their Hubble diagram is plotted in a Local-Group-centric reference frame in \cref{fig:eddall}, while
their on-sky distribution is shown in \cref{fig:eddsky}. We note that the spatial distribution of galaxies more distant than \SI{2}{Mpc} follows the Supergalactic plane quite closely.
We note that because our likelihood is conditioned on the flow velocity at the position of each galaxy, we should not be biased due to the inhomogeneity and incompleteness of the sky distribution of our sample. 
Analyses that assume spherical symmetry of the tracer populations and of the surrounding Hubble flow when determining quantities like the Local Group mass and turnaround radius may be biased by such issues \citep[see also][]{santos-santosAnisotropiesSpatialDistribution2023}.

\subsubsection{Correcting masses for resolution effects}
\label{sec:masscorrectionmodel}
For our purposes, the main effect of low numerical resolution in our simulations is that haloes become less concentrated so that the mass within a Gaussian filter is biased low. This bias is quite systematic and can be well described by a simple power law: $M_\text{gadget} = A M_\text{\borg low-res} ^ \beta$, with some log-normal scatter $\sigma_{\ln{M}}$.
Hence, if we identify simulation parameters for which $\sigma_{\ln{M}}$ is small, we can adequately correct for the effect of lowered resolution.
To this end, we ran unconstrained simulations similar to the one shown in \cref{app:masscorrection}, but with varying spatial, time and mass resolution. We then cross-matched individual haloes in the \borg and Gadget simulations, and fit power-laws linking the halo masses in the two schemes, determining the mean relation and the scatter for these cross-matched haloes.

\subsubsection{Velocity reference frame}
\label{sec:veloffsetframe}
To compare observational data to our simulations using the velocity likelihoods above, we had to find the velocity offset between the real heliocentric frame and the simulation reference frame. 
This is done by assuming that the real Galactocentric frame coincides with the simulation reference frame in which the simulated Milky Way has a zero \SI{100}{kpc} filtered velocity. 
However, in reality, there might be a difference between the \SI{100}{kpc} filtered velocity and the velocity of the Galactic centre, and additionally, there is some uncertainty in the Solar Galactocentric velocity vector.
We incorporate these uncertainties by marginalising over a possible offset velocity, $\vb*{v}_\text{off}$ which we assume to have zero mean and a \SI{10}{km.s^{-1}} isotropic scatter (see \cref{app:marginalisation}).
Since all our velocity likelihoods are multivariate normal distributions, this marginalisation can be done analytically. The derivation is shown in \cref{app:marginalisation}. 

\section{Results}
\label{sec:results}
To present our results, we first describe the experimentation and reasoning underlying the particular simulation parameters we end up using. (\cref{sec:simpars}). We then present some results derived from the analysis of our chains in \cref{sec:chains}, we quantify the resulting autocorrelation in \cref{sec:autocorrelation}, and we provide some noteworthy quantitative predictions in \cref{sec:posteriors} and in \cref{sec:masslocalgroup}.

\subsection{Simulation parameters}
\label{sec:simpars}
Our set-up incorporates constraints from different aspects: model uncertainties, numerical accuracy, and resolution. We summarise here the considerations for each parameter.

In the first place, there is the log-likelihood itself, which we would like to represent the data as closely as possible. Ideally, we would like to match the positions of the Milky Way and M31, and their relative radial velocity to within the observational uncertainties. However, the measurement errors for these data are very small. This leads to large numerical inaccuracies in the HMC algorithm, in turn requiring very small timesteps in order to reach acceptable acceptance rates \citep[we aim for \SIrange{50}{70}{\percent}, which is optimal for Gaussian distributions,][]{brooksHandbookMarkovChain2011},
resulting in greater autocorrelation lengths, and so substantially longer computation time to reach the same effective number of samples. To overcome this problem, in \cref{sec:likmass,sec:likvelocity} we purposely degraded the precision of the constraints on halo positions and relative radial velocity to a level that is compatible with the accuracy of our simulation, taking into account its resolution and the lack of baryonic effects.

To tune the resolution and the particle masses, we require that simulations at high resolution carried out with Gadget match lower resolution runs carried out with our Zoom algorithm. In practice, we found that our filtered halo masses are most sensitive to lowering resolution. Therefore, we tune our resolution parameters to minimise runtime, while still being able to correct for the effect of low resolution on the masses. In practice, we choose parameters for which the uncertainty in the halo mass bias correction $\sigma$ is below 10\% (see \cref{app:masscorrection} for the size and uncertainty of the correction for our preferred parameters).

We note that we verify in \cref{sec:resimulations} that for our chosen spatial, mass and time resolutions, our final masses, positions and velocities are sufficiently accurate for our problem. The choice of resolution has multiple consequences and it is important to limit it for two reasons: higher resolution leads to more costly computations, but it also leads to a rougher probability landscape which impacts negatively the statistical performance of the HMC.
One possible explanation for this might be that higher resolution in space and time allows for more complicated particle trajectories which are then less linear, making the problem harder to sample from. 

The sizes of the low- and high-resolution regions were determined by requiring the outer box to be large enough to be able to fit the local Hubble flow while being small enough that the Lagrangian region at its centre does not move substantially. 
If the bulk flow were too large compared to the size of the high-resolution box, there would be substantial leakage of low-resolution particles, or of particles that have a major part of their history inside the low-resolution region, into the region of interest (the MW-M31 pair and its immediate surroundings).
The high-resolution region must be large enough to allow some flexibility, yet small enough that a computationally affordable grid ($128^3$ is currently feasible) can resolve the formation of the MW and M31 haloes.

The initial Gaussian white noise fields themselves are on a $64^3$ grid for the LR field and a $80^3$ grid for the HR field, due to the need for zero-padding. This gives our inference problem a total dimensionality of $N_\text{dim}=64^3+80^3=774~144$. The particles themselves are placed at the centres of the voxels of two $64^3$ grids. The forces are calculated on grids of twice this size, i.e. a $128^3$ grid and a $160^3$ grid respectively. 

To integrate the equations of motion of our N-body simulations, we use the default scheme in \borg \citep{jaschePhysicalBayesianModelling2019} for which timesteps are linearly spaced in the cosmological scale factor $a$.

The resulting numerical parameters are summarised in \cref{tab:simsettings}. With these settings, we then proceed to sample from our initial condition field $\vb*s$.

\begin{table}
\centering
\begin{tabular}{lrr}
\toprule
& Low-res region & High-res region\\
\midrule
Box size $L$ & \SI{40}{Mpc} & \SI{10}{Mpc} \\
Force mesh voxel size & \SI{312.5}{kpc} & \SI{78.125}{kpc}\\
Particle mass & \SI{9.65e9}{\Msun} & \SI{1.51e8}{\Msun} \\
Initial condition field size & $64^3$ & $80^3$ \\
Starting redshift & 63 & 63 \\
Number of timesteps & 40 & 40 \\
\midrule
HMC step size & \multicolumn{2}{c}{\num{1e-3}} \\
HMC number of steps & \multicolumn{2}{c}{200} \\
\bottomrule
\end{tabular}
\caption{A summary of the simulation settings. 
The particle mass refers to the mass at the initial time, at later times, there will be some leaking of low-resolution particles into the high-res region and vice versa.\label{tab:simsettings}}

\end{table}

\subsection{Markov chains}
\label{sec:chains}

\begin{figure*}
    \centering
    \includegraphics[width=\linewidth]{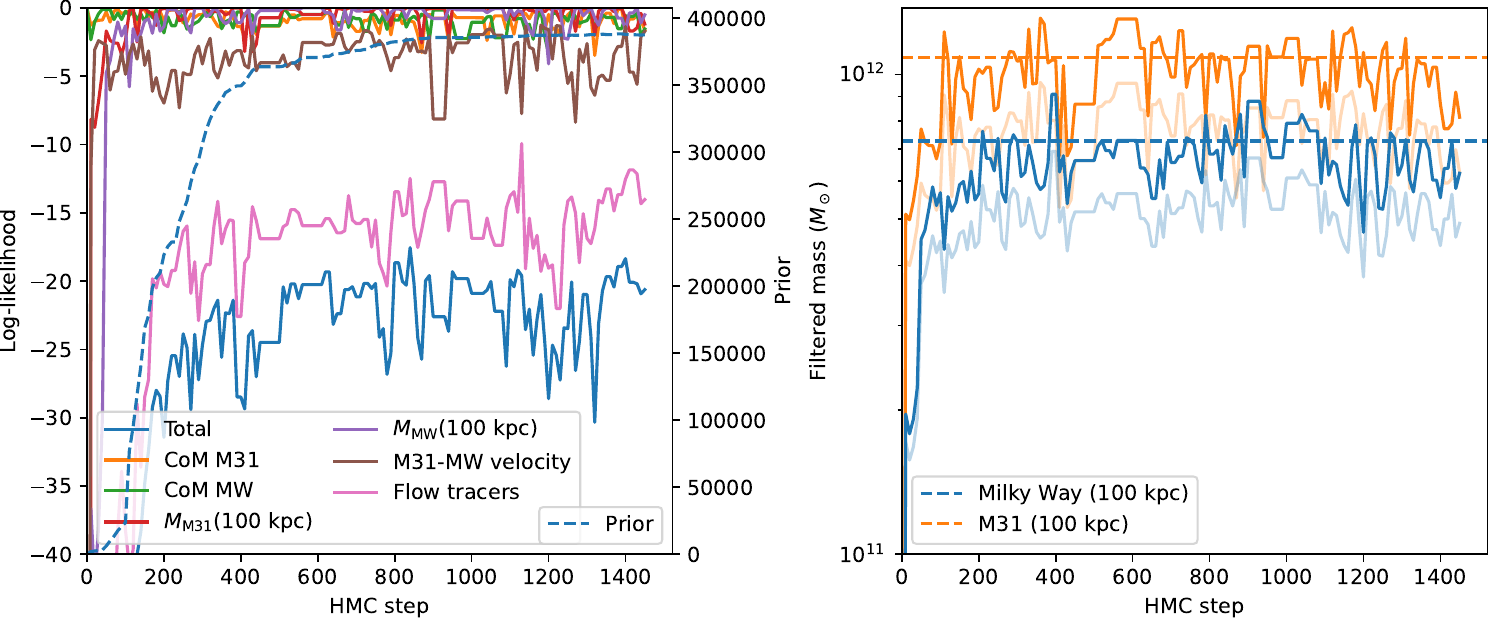}
    \caption{Left: the likelihood values of the observational constraints of an example chain, chain number 10. The solid blue line is the total likelihood, which is the sum of all other solid lines that indicate components of the likelihood. The dashed lines are the prior on the initial white noise fields $\vb*s_\text{LR}$ and $\vb*s_\text{HR}$. Right: The masses of the Milky Way and M31 along the chain. The greyed-out lines indicate values of the masses before applying the correction from \cref{fig:masscorrection}. This particular chain was considered warmed up by sample 760 according to the criterion of \cref{sec:autocorrelation}.}
    \label{fig:ch6}
\end{figure*}

For the work presented in this paper, we have run in parallel 12 Markov chains. A trace plot of the different contributions to the posterior of one HMC chain is shown in the left panel of \cref{fig:ch6}.
In this plot, the different terms that make up the likelihood are shown with different colours, namely the log-likelihood of the individual halo masses (red and purple, \cref{eq:likMMW,eq:likMM31}) and positions (orange and green, \cref{eq:likcommw,eq:likcomm31}) for a \SI{100}{kpc} filter, the 3D M31-MW relative velocity (brown, \cref{eq:vdiffm31mw}), and the total likelihood of the Hubble flow tracers (pink, \cref{eq:vhelio}). After burn-in these are expected to vary like a chi-square distribution (scaled by $-1/2$) with one degree of freedom (masses), 3 degrees of freedom (positions and 3d-velocity), or 31 degrees of freedom (flow tracers, since we include 31 galaxies).

Also shown is the value of the prior log-likelihood (dashed blue and dashed orange), the $\chi^2/2$ of the initial Gaussian random fields that generate the initial conditions.
During warm-up, the Gaussian random numbers that generate the initial conditions were set to start at zero for all voxels. This procedure is commonly used within \borg to speed up the warm-up process; we aim to improve it in future. Since this prior is just the sum of squares of Gaussian random numbers, after warm-up its value should follow a chi-square distribution (up to a division by two) hence have mean $N_\text{dim}/2$, where, in our case, $N_\text{dim}/2=\num{387072}$.
That is why the dashed lines in \cref{fig:ch6}, which indicate the value of the prior, start at $0$ and then converge around $387~000$.

\cref{fig:ch6} shows that a hierarchy exists, where small-scale constraints like the individual halo properties warm up very quickly, followed by constraints on larger scales such as the MW-M31 velocity (brown) the local Hubble flow (pink), and finally the field as a whole (the dashed lines showing the prior). All the observational constraints get to within a few $\sigma$ of their required observational values by about 200 HMC steps, the last being the flow tracer velocities (pink curve) which should scatter within a scaled $\chi^2$-distribution with a mean value half the number of galaxies we consider, so $31/2=15.5$. 

In the right panel of \cref{fig:ch6}, the simulation-derived masses are shown.
We reach convergence within a few hundred samples, and the masses for both the Milky Way and M31 are biased slightly low, even after applying the \cref{app:masscorrection} correction which accounts both for the low spatial and time resolution of the simulation, and for the fact that we determine masses for fixed filter positions which do not coincide exactly with the centres of the simulated haloes. 
In principle, one could avoid the problem of haloes being offset from the filter centre by first ``finding" the centre of the halo at each step, but this is not straightforward: one would have to make sure that the procedure stays differentiable, and also that the centre of mass varies continuously as a function of initial conditions. Failing to require these two properties would cause the log-likelihood to be discontinuous, resulting in poor sampling performance.

\subsection{Autocorrelation analysis}
\label{sec:autocorrelation}
\begin{figure*}
    \centering
\includegraphics[width=\linewidth]{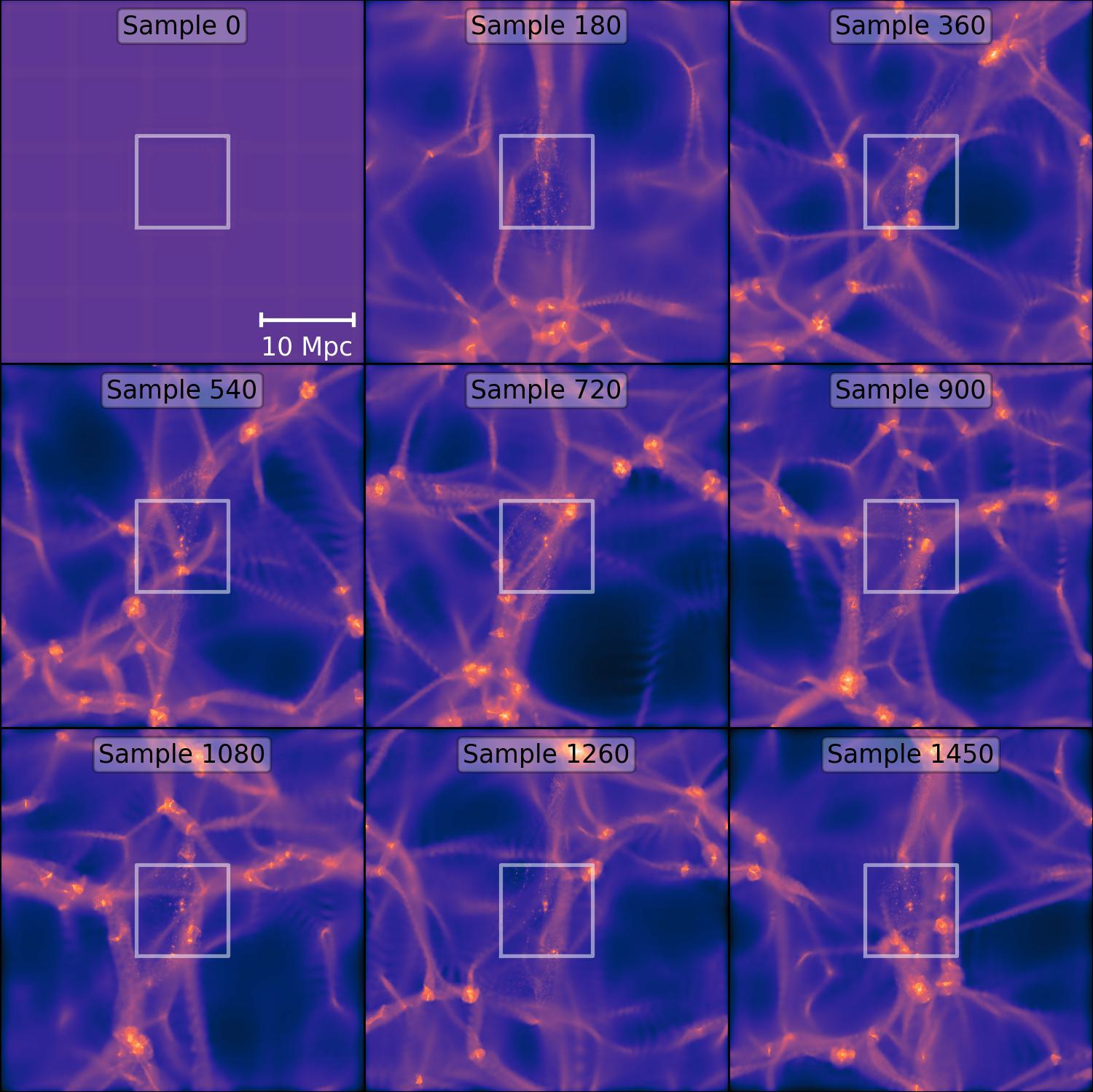}
    \caption{The final density field of the low-resolution box as chain 10 evolves. This chain has $F_\text{eff}=435$, so every $F_\text{eff}$ sample can be considered as an independent sample. Note that sample $0$ (which occurs after one iteration of the HMC) is almost completely uniform because we start our chains from a uniform field. Visualisation was done using \texttt{pySPHViewer} \citep{PySPHViewerV12015}.}
    \label{fig:ch844bigbox}
\end{figure*}

In \cref{fig:ch844bigbox} we show images of the final state of nine simulations spaced evenly along the chain of \cref{fig:ch6}.
The present-day density field reflects directly the initial conditions, and one can see correlations in structure between neighboring images in \cref{fig:ch844bigbox}; present-day density fields (and therefore the corresponding initial conditions) do not fully decorrelate between samples that are not spaced sufficiently far apart. For instance, in consecutive images, one can see large haloes repeating, shifted only by relatively small amounts. Some coherence between nearby samples is also visible in the behaviour of the flow tracer likelihood (the pink curve in \cref{fig:ch6}).

\begin{figure}[tpb]
    \centering
    \includegraphics[width=\linewidth]{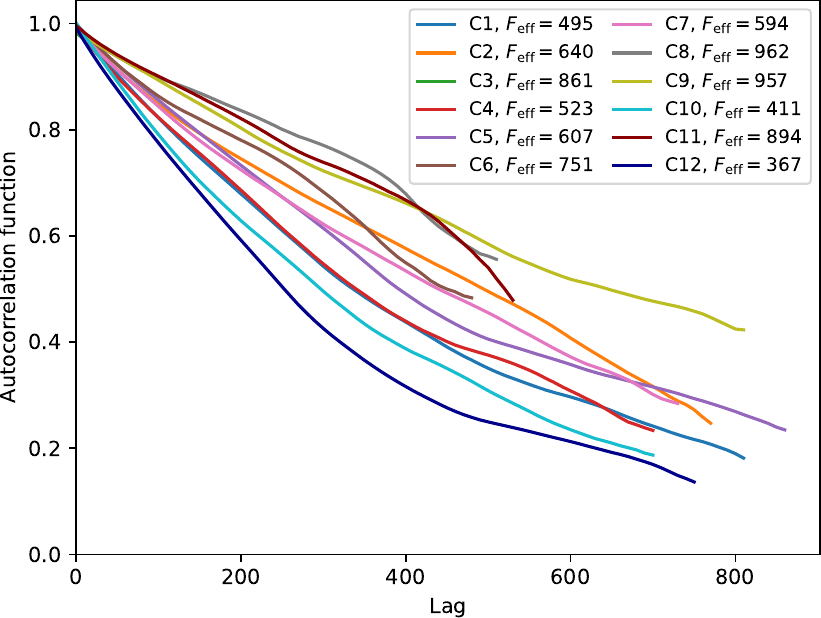}
    \caption{The autocorrelation functions as computed using \cref{eq:autocorrelation} for our twelve different chains, each indicated by a different colour.}
    \label{fig:autocorrelations}
\end{figure}
It is helpful to quantify correlation as a function of lag within a chain, which
we choose to do by computing correlations between initial condition fields $\vb*{s}$. These decorrelate the slowest among all the quantities we have considered, so this is a conservative choice. 
We compute the autocorrelation along the chain of the vector $\vb*{s}_i$, representing both the HR and LR initial conditions, which we calculate for a lag $t$ as 
\begin{equation}
    \rho_t = \frac{1}{N_\text{dim}}\frac{1}{N - t}\sum_{i=1}^{N-t}{\vb*{s}_i^T\vb*{s}_{i+t}}, \label{eq:autocorrelation}
\end{equation}
which is equivalent to the average fictitious time autocorrelation of the individual components of $\vb*s$. We use only samples after the chains are warmed up.
Since the effective sample size is \citep{brooksHandbookMarkovChain2011}
\begin{equation}
    N_\text{eff} = \frac{N}{\sum_{t=-\infty}^\infty \rho_t},
\end{equation}
we can find the effective sample size through an estimate of $\rho_t$.
One would ideally use directly estimated values for $\rho_t$; unfortunately, if $t$ is too large, we are limited by the length of the chain.
In practice, $\rho_t$ has exponential behaviour, which is common for HMC where each step has a small integration length. Thus, for each chain, we fit an exponential curve, and from this fit, we can obtain a good estimate of $\rho_t$ for all $t$. We show the correlation functions $\rho_t$ for our different chains in \cref{fig:autocorrelations}, giving also the $N_\text{eff}$ derived from exponential fits.

The number of samples per effective sample, $F_\text{eff}$, is
\begin{equation}
F_\text{eff} = \sum_{t=-\infty}^\infty \rho_t, \label{eq:thinningfactor}
\end{equation}
so to obtain independent samples, we can thin each chain with the corresponding $F_\text{eff}$, taking samples separated by $F_\text{eff}$, starting from the last sample in the chain. Proceeding in this way, we obtain $27$ independent samples across all chains, four of which are shown in \cref{fig:visualisation}. In \cref{fig:ch844bigbox}, an example chain is shown, where one can verify by eye that samples separated by $F_\text{eff}=411$  appear fully decorrelated. This decorrelation number varies from chain to chain as can be seen seen in \cref{fig:autocorrelations}. 

Since the initial condition field decorrelates the slowest, we can use this to provide a simple criterion to indicate when the chain has warmed up. If the prior, (which started at zero) is above \SI{98}{\percent} of its final value (this means that the power spectrum is within a few per cent of the cosmological expectation) we consider the chain to be sufficiently warmed-up.

For this paper, we create two sets of samples:
\begin{itemize}
    \item An ``independent" set of simulations, where each chain has been thinned by the factor $F_\text{eff}$ above, i.e. first taking the sample at the end of the chain, then the sample $F_\text{eff}$ before that, then repeatedly stepping back by $F_\text{eff}$ until we reach the warm-up phase.
    \item A ``semi-independent" set of simulations, which is the same, but thinned by a factor $F_\text{eff}/20$ rather than $F_\text{eff}$. Despite large scales being more correlated in nearby members of this set, quantities determined primarily by small-scale power, such as individual halo masses, are largely independent.
\end{itemize}

We believe that we are rather conservative in our estimate of the true effective number of samples: when visually inspecting a chain, one can see almost no correlation when considering two samples separated by $F_\text{eff}$. We note that the inter-chain variation is similar to the intra-chain variation.

We note that, in any MCMC analysis, the posterior mean of any inferred quantity $q$ has a corresponding error due to Monte Carlo integration. This error is $\sim\!\sigma(q) / \sqrt{N_\text{eff}}$  \citep{brooksHandbookMarkovChain2011}. In practice, this error for $N_\text{eff}=27$, is of the order of $\SI{19}{\percent}$. Adding in quadrature to the full posterior width $\sigma(q)$ gives a total error of $\sqrt{\sigma(q)^2+(0.19\sigma(q))^2}=1.02\sigma(q)$, a change of \SI{2}{\percent} which barely changes the resulting uncertainty. A typical case in this work is the estimation of masses. 

\subsection{Posterior quantities for the individual galaxies}
\label{sec:posteriors}
\begin{figure*}
{
    \centering
    \includegraphics[width=0.49\linewidth]{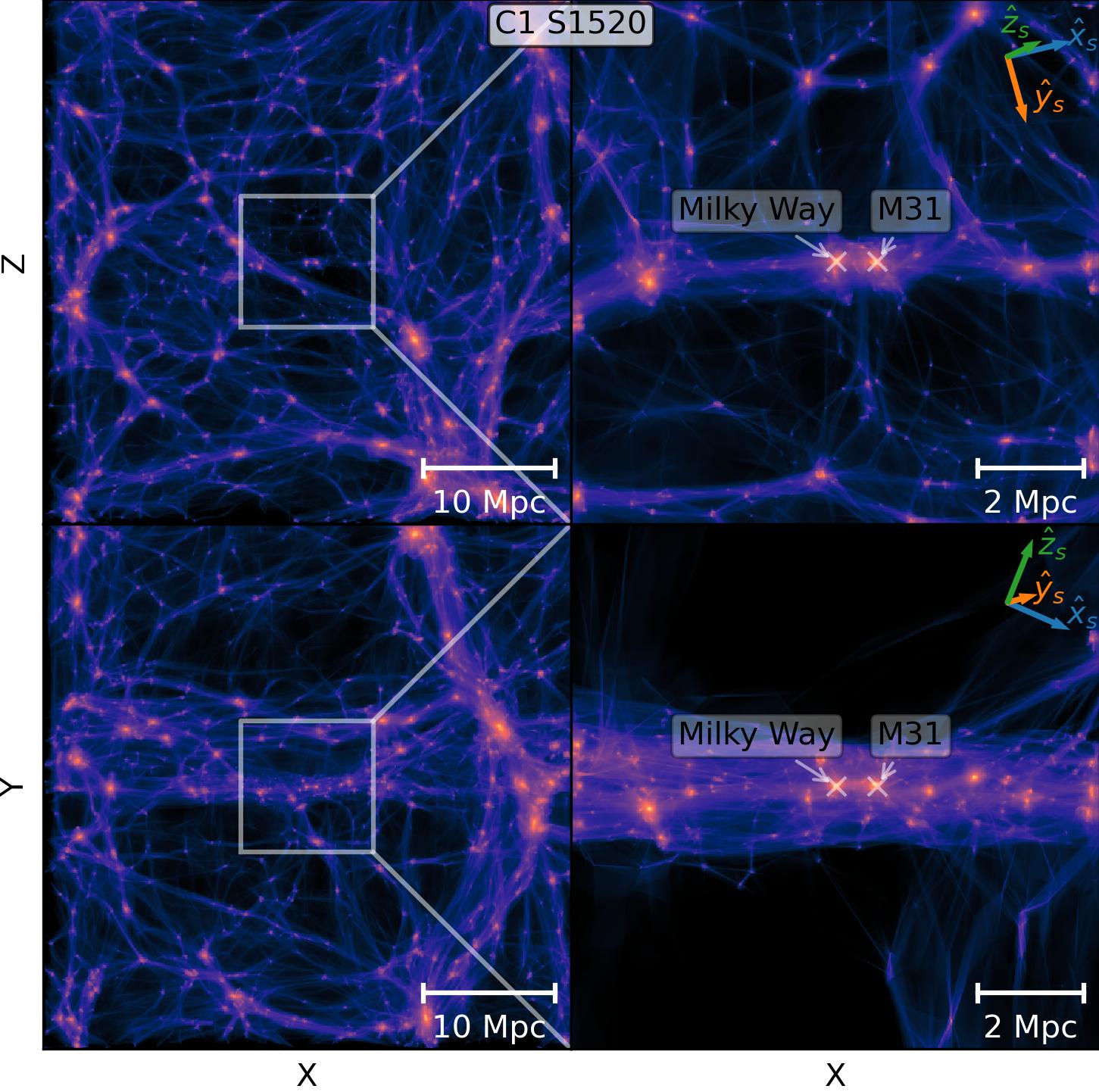}
    \includegraphics[width=0.49\linewidth]{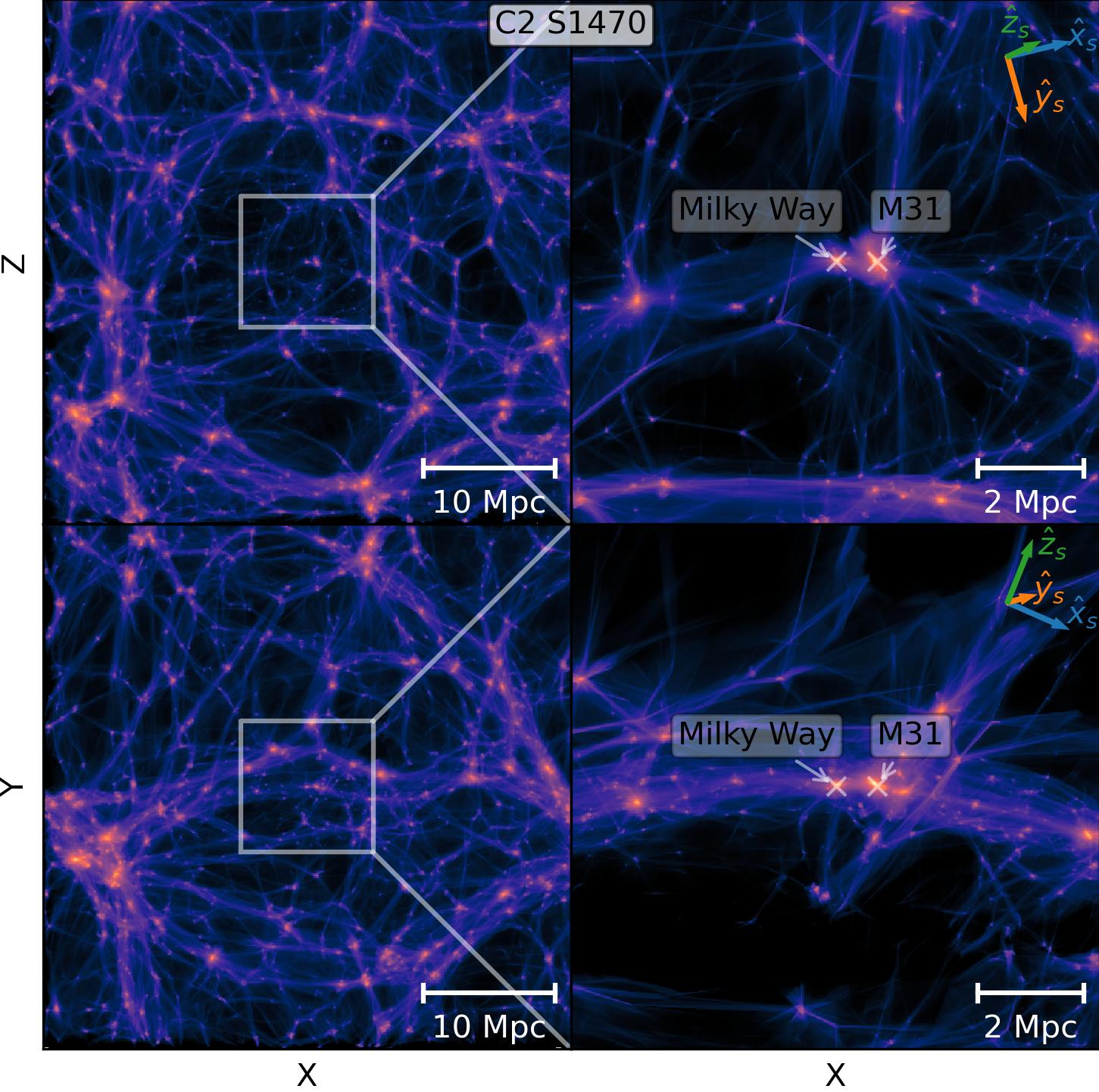}
    \includegraphics[width=0.49\linewidth]{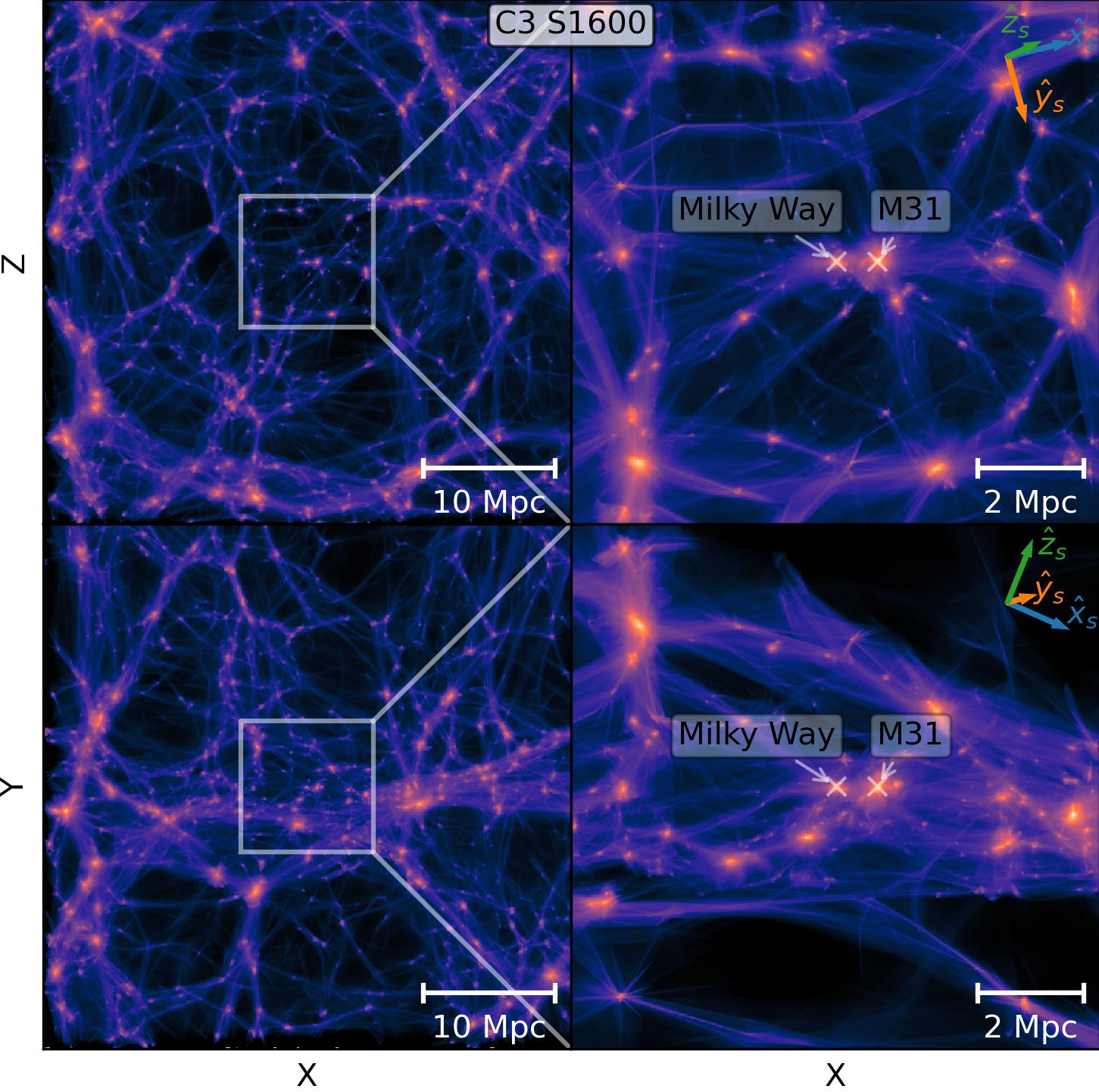}
    \includegraphics[width=0.49\linewidth]{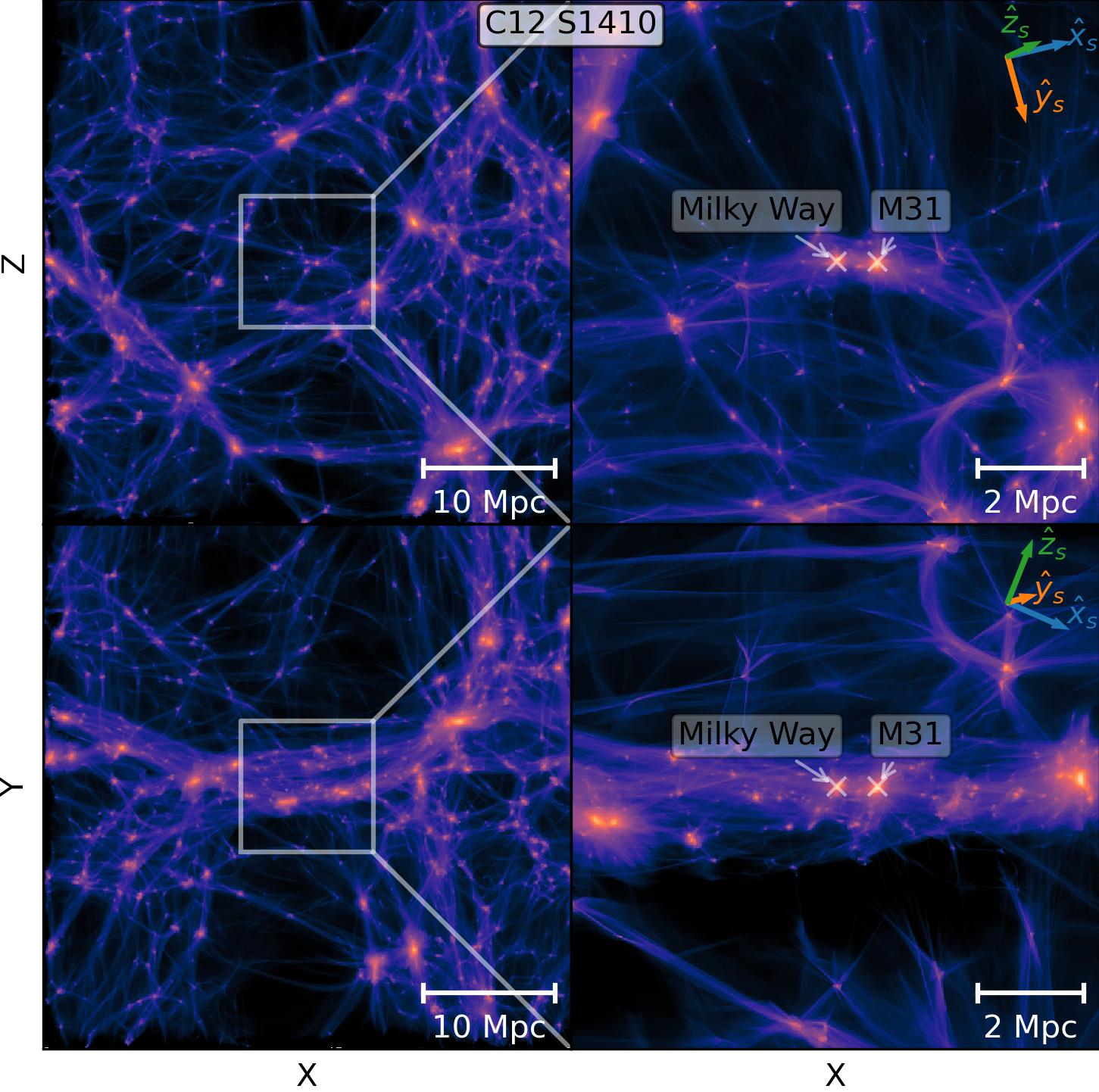}
    \caption{The final density fields in simulation coordinates of evolution from four of our independent sets of initial conditions (each from a different chain).
    Each pane corresponds to a single set of initial conditions. Within each pane, the left column shows the full box and the right column the central \SI{10}{Mpc} zoom box, while the two rows show orthogonal projections.
    The name at the top of each pane indicates the chain it came from (after \emph{C}), and its sample index (after \emph{S}).
    The visualisations are made using the Lagrangian Sheet density estimation method of the \emph{r3d} package\textsuperscript{*} \citep{powellExactGeneralRemeshing2015}. The unit vectors of the Supergalactic reference frame are shown at the top right of each high-resolution panel.
    }
    \label{fig:visualisation}
    }
    \small\textsuperscript{*} Since our two-grid initial condition layout is not a pure cubic grid, in particular, at the zoom-region boundary, we cannot use the recipe where every cube is divided into 6 tetrahedra as in e.g. \citet{abelTracingDarkMatter2012}. Instead, we use a Delaunay tetrahedralisation to create the simplex tracers.
\end{figure*}
\begin{figure*}
    \centering
    \includegraphics[width=\linewidth]{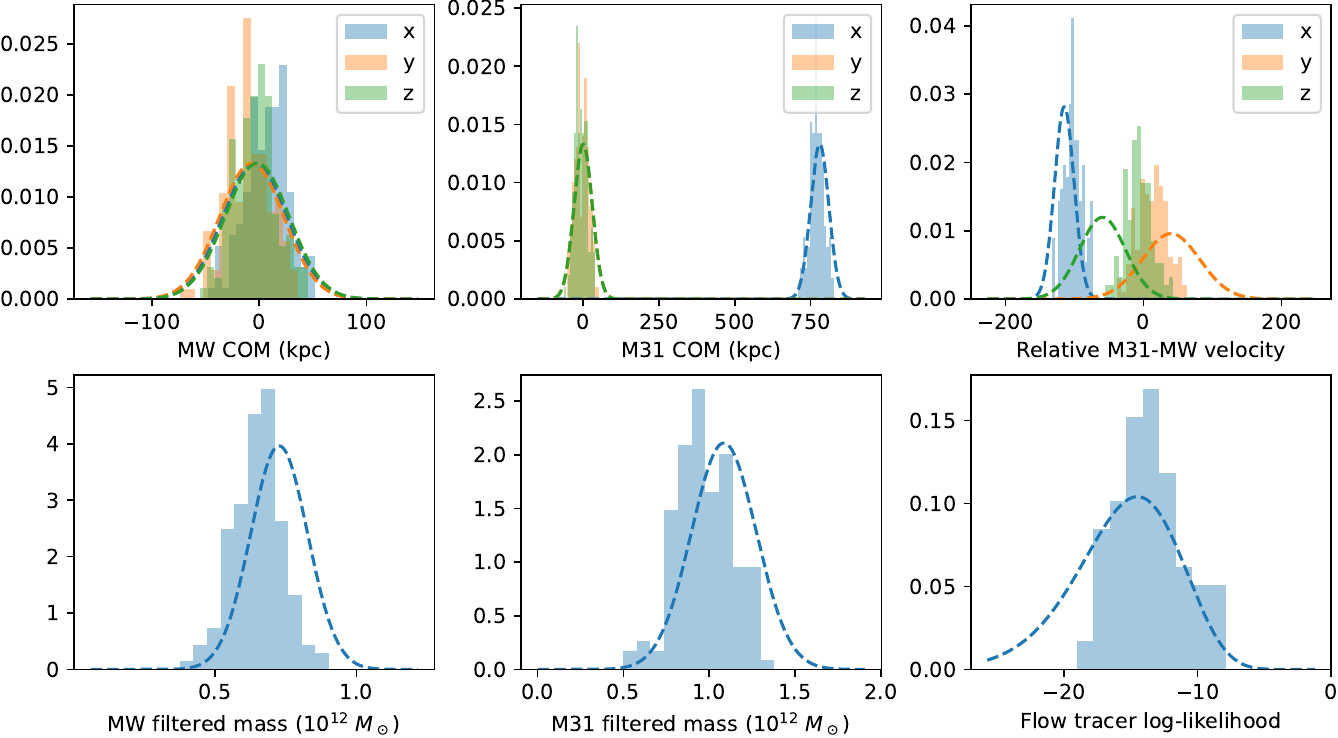}
    \caption{The posterior distributions of Local Group properties from our Markov chains. In each panel, the histogram is of samples from our chains and therefore indicates the posterior distribution. The dashed lines are the injected observational constraints. The top left panel indicates the centre-of-mass position of the Milky Way, and the top middle panel M31. The right panel shows the relative velocity of the MW-M31 pair. The bottom left and middle panels indicate the \SI{100}{kpc} filtered masses of the Milky Way and M31. The bottom right panel shows the total log-likelihood of the flow tracers, with on top the $\chi^2$-distribution (scaled by $-\frac{1}{2}$, since the log-likelihood is $-\chi^2/2$). These histograms are of the ``semi-independent" set of simulations as described in \cref{sec:autocorrelation}.}
    \label{fig:posteriors}
\end{figure*}
With a reasonable set of independent samples, we can make a variety of quantitative statements based on our chains. A few examples of ``independent" samples of the Local Group environment (each from a different chain) are shown in \cref{fig:visualisation}.
In every sample, the Milky Way and M31 lie within a connecting filament. On a larger scale, the Local Group lives in a relatively quiet, underdense, environment, in a wall-like structure that is approximately aligned with the Local Sheet: most structures live close to the $xy$-plane in Supergalactic Coordinates. This is {\it a priori} somewhat surprising, since we have not explicitly constrained the density field surrounding the Local Group, only the peculiar velocity field, which appears close to isotropic.

In \cref{fig:posteriors}, we show posterior distributions for various quantities along our chains. 
These histograms indicate plausible values for quantities like the filtered halo masses of the Milky Way and Andromeda, given our full set of constraints and our specific assumed \LCDM model, whereas the dashed lines show the observationally-based constraints that we adopted as priors for each quantity, as described in \cref{sec:lglik}; for the filtered halo masses, these are already the result of Appendix \ref{app:observationalmasses}'s quasi-Bayesian analysis of observational data for halo mass tracers surrounding the two galaxies under a \LCDM prior. 

The centre-of-mass distributions of the two main haloes match well with the likelihood constraints we have put on their positions. In fact, they are less tail-heavy than the input, which is expected, because moving the centre of mass too far away would also impact other terms in the likelihood, in particular, the mass likelihood.

While the \SI{100}{kpc}-filtered M31 mass matches the injected observational constraint quite well, with only a small downward bias, the filtered Milky Way mass is biased more substantially  low. These biases are however mostly a consequence of the halo centres not aligning perfectly with the filter centres; the biases decrease to within a few percent of the observational prior when re-centering the filters at the precise halo centre. Our chains imply virial mass estimates for the two haloes of $\log_{10}(M_{200c}/M_\odot) = 12.07\pm0.08$ and $12.33\pm0.10$, for the MW and M31, respectively.\footnote{This is defined as the mass within a sphere that has an average density of 200 times the critical density of the universe. The centre of the sphere is chosen to be the shrinking-spheres centre. We correct for the effect of limited resolution inside \borg using the correction for $M_\text{200c}$ in \cref{app:masscorrection}.} This is very similar to the implied $M_\text{200c}$'s we obtain from our NFW-like fitting to observation in \cref{app:observationalmasses}.  Note the small error bars resulting from the need to satisfy all observational constraints simultaneously. The fact that these masses are near the lower end of the range quoted in earlier studies is mainly a consequence of the Hubble flow constraint, which not only requires a relatively small total mass for the Local Group but also approximately determines its centre of mass, thereby implying a mass ratio for the two main galaxies \citep[see][]{penarrubiaDynamicalModelLocal2014}. 
With their simple spherical model, these authors found a mass ratio of $0.54\substack{+0.23 \\ -0.17}$ (mean and 16th/84th percentiles) whereas we find a mass ratio of $M_\text{200c,MW}/M_\text{200c,M31} = 0.57\substack{+0.12 \\ -0.16}$, similar to the value $0.58\pm 0.19$ implied for this ratio by the analysis of \cref{app:observationalmasses}. These results are discussed further in \cref{sec:massdistributionlgdiscussion} below.

\subsection{The environment and mass of the Local Group}
\label{sec:masslocalgroup}
\begin{figure}
    \centering
    \includegraphics[width=\linewidth]{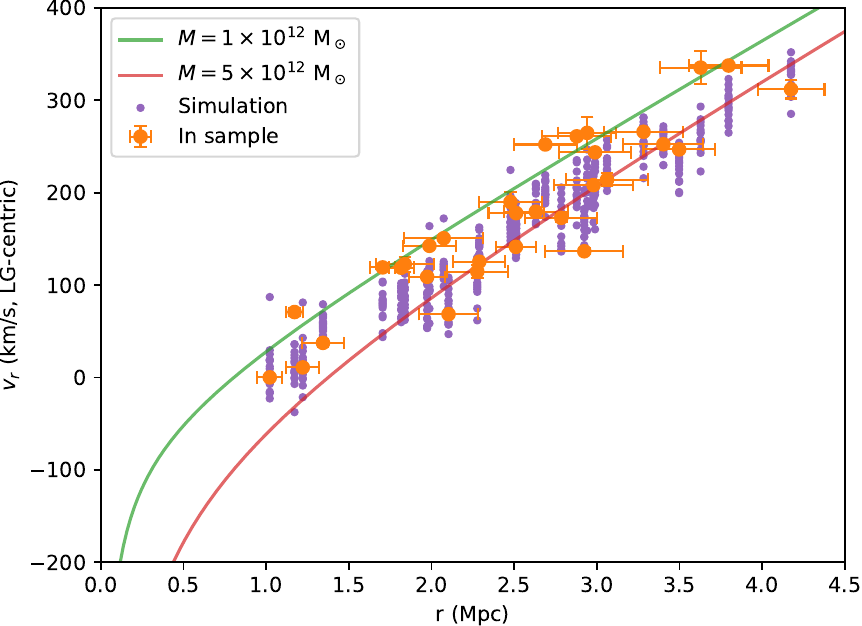}
    \caption{Similar to \cref{fig:eddall}, but now with the purple points showing the line-of-sight velocities from our simulations at the 3D location of each galaxy.
    Each purple point is an independent sample from our chains, specifically, we plot the ``independent" set as defined in \cref{sec:autocorrelation}.}
    \label{fig:hubblesim}
\end{figure}

Visual inspection of our chains shows that they all produce first-approach trajectories for the MW-M31 pair. Furthermore, orbits for the pair are almost perfectly radial, i.e. along $v_x$. The tangential components in the $(\alpha, \delta)$ directions are $(v_\alpha, v_\delta) = (\num{12.7(213)}, \num{-6.7(187)})\ \si{km.s^{-1}}$, while the velocity in the radial direction has a posterior of $v_x=\SI{-102(13)}{km.s^{-1}}$.
The small tangential motion is likely a result of the constraints on the local Hubble flow that we have applied because a significantly non-radial orbit would require a nearby massive object in order to produce sufficient tidal torque to generate the orbital angular momentum. The presence of such an object is disfavored by the quiet and at most weakly distorted Hubble expansion indicated by our flow tracers. We can see in \cref{fig:hubblesim} that the velocities of our isolated galaxy catalogue are generally well matched by our simulations, with one or two possible exceptions (e.g. Tucana) which would be worth more detailed investigation. 

\begin{figure}
    \centering
    \includegraphics[width=\linewidth]{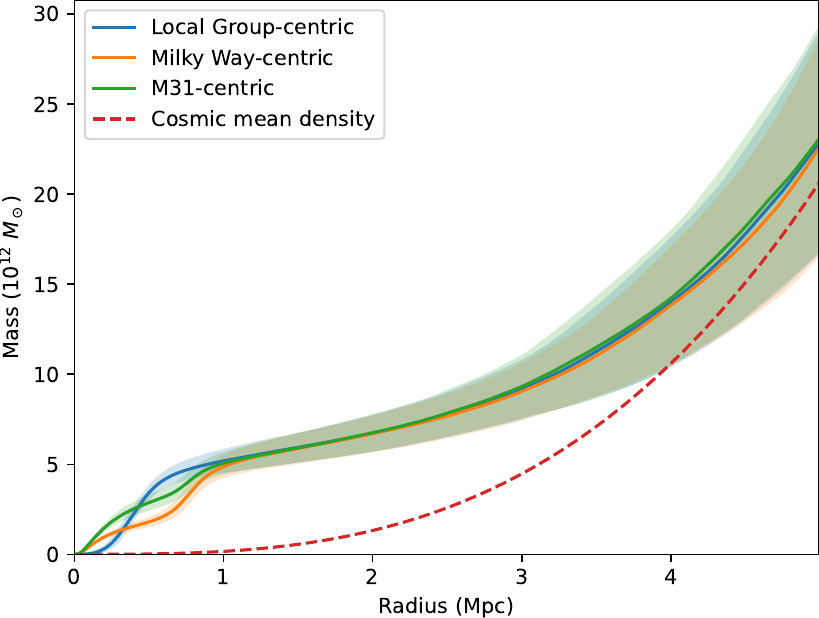}
    \caption{Enclosed mass profiles in our simulations, centred on the Local Group centre-of-mass, on the Milky Way, and on M31. The lines indicate the posterior mean, and the shaded regions are the $1\sigma$ region of the ``semi-independent" sample set (see \cref{sec:autocorrelation}, we use this to avoid being affected by shot noise coming from thinning the chains).}
    \label{fig:massprofiles}
\end{figure}

Inference on the mass profiles is shown in \cref{fig:massprofiles}. We find posterior distributions for the enclosed Local Group mass, $M(\SI{<1}{Mpc}) = \SI{5.20(67)e12}{\Msun}$ and $M(\SI{<2.5}{Mpc}) = \SI{7.77(127)e12}{\Msun}$, when measured from the Local Group barycentre\footnote{In this case, the barycentre is found by finding the centre-of-mass of all particles within a \SI{1}{Mpc} sphere centred at the point between the Milky Way and M31, although in practice the masses are rather insensitive to the exact chosen centre when considering radii beyond \SI{1}{Mpc}.}. 
Within the larger radius, the Local Group is only overdense by a factor of about \num{3.00(49)}. By a radius of \SI{4}{Mpc}, the distance of the most distant flow tracers we have used, this factor has dropped to \num{1.33(35)}.
If the Local Group mass is instead defined as the sum of the virial masses $M_{200c}$ of the two objects, we find $M_\text{LG}= \SI[parse-numbers=false]{3.34\substack{+0.58 \\ -0.54}\times10^{12}}{\Msun}$.
It is interesting to compare these numbers to previous work. From the Timing Argument alone, \citet{liMassesLocalGroup2008} found the sum of the two $M_{200c}$ values to be $\log_{10}(M_\text{LG}/\si{\Msun})= 12.74\substack{+0.17 \\ -0.27} $ for a \LCDM prior. 
This is substantially larger than our value (although consistent with it at about 1$\sigma$) reflecting the fact that the quiet local Hubble flow is not consistent with the upper end of their Timing Argument range.
In contrast, using local Hubble flow observations alone, \citet{penarrubiaDynamicalModelLocal2014} estimated the mass of the LG within about \SI{0.8}{Mpc} to be $M(\SI{<0.8}{Mpc}) = \SI{2.3(7)e12}{\Msun}$, consistent only at about 2.8$\sigma$ with the mass, $M(\SI{<0.8}{\Msun}) = \SI{4.82(58)e12}{\Msun}$, that we find; such a low mass is strongly disfavored by the Timing Argument. Combining the two types of data, i.e. MW+M31's orbital/halo properties together with the quiet Hubble flow, leads to our own tighter constraints.

Finally, in the bottom right panel of \cref{fig:posteriors}, we see that the distribution of total log-likelihood values for galaxy velocities in the Local Group's immediate environment is consistent with the expected $\chi^2$-distribution. Since the uncertainties entering this estimate are dominated by the \SI{35}{km.s^{-1}} residual scatter assumed in our likelihood, the fact that we obtain fully acceptable $\chi^2$ values means that our \LCDM simulations match the observational data as well as the simple spherical model of \citet{penarrubiaDynamicalModelLocal2014}.


\section{Resimulations with Gadget-4}
\label{sec:resimulations}
The simulations discussed so far are based on a zoom scheme with a relatively small grid size for the two nested meshes. This is optimised for computational speed but is relatively low-resolution. In particular, the spatial and time resolution are sub-optimal for detailed studies of the structure and evolution of the Local Group galaxies. Thus, it is necessary to see how well the initial conditions identified by our procedure perform when the resolution in space and time is improved and high-resolution simulations are carried out with state-of-the-art software. 
Here we compare with Gadget-4 \citep{springelSimulatingCosmicStructure2021} in a simple Tree-PM setting (non-zoom) but with the same initial particle masses, positions and velocities as in the corresponding \borg simulation. Hence, these runs test only the effect of reducing the timesteps and the softening to values comparable to those usually used in high-resolution calculations.
The (comoving) force softenings that we adopt are \SI{6}{kpc/h} for HR-particles and \SI{24}{kpc/h} for LR-particles. This approximately follows the rule of thumb proposed by \citet{powerInnerStructureLCDM2003}, namely $\epsilon_\text{opt} \approx 4 r_{200}/\sqrt{N_{200}}$, which for a $\SI{1.5e12}{\Msun}$ halo with virial radius \SI{\sim250}{kpc} and particle mass \SI{1.5e8}{\Msun} leads to \SI{\sim10}{kpc}.
Using the initial condition fields $\vb*s$ that we found from our chains, we generate the particle's initial positions and velocities using the technique described in \cref{sec:methods}, specifically \cref{eq:zoomlpt}.
We put into Gadget the particle masses, positions and velocities that the original \borg scheme generated at its starting redshift, $z=63$.

\begin{figure*}[htpb]
    \centering
    \includegraphics[width=\linewidth]{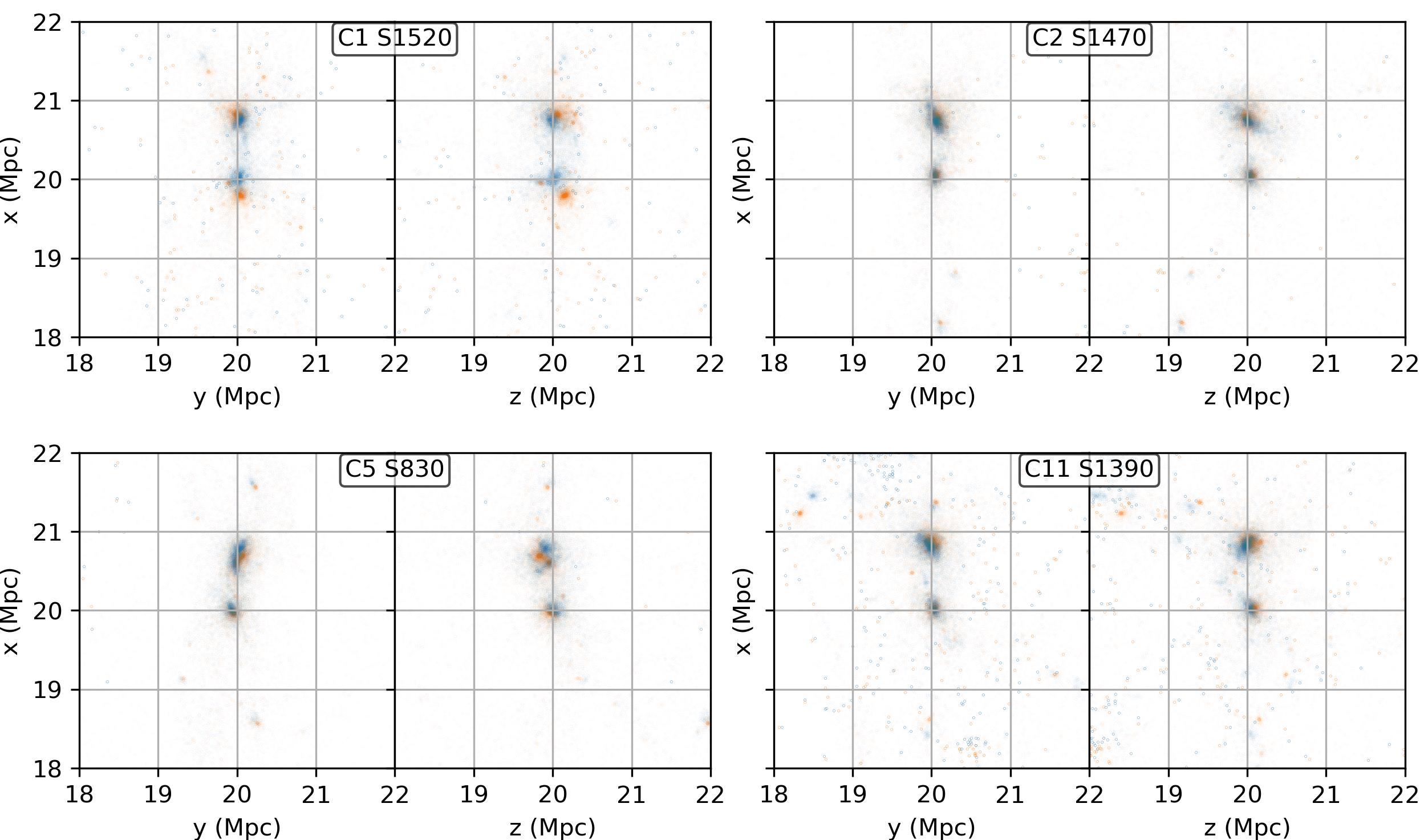}
    \caption{The Local Group region (a box of $(\SI{2}{Mpc})^3$ centred at the Milky Way) in some resimulations with Gadget-4. The blue dots indicate particles in the \borg simulations, and the orange dots indicate particles in the gadget re-simulations. Each panel indicates one simulation, with two projections being shown.}
    \label{fig:gadget_resims}
\end{figure*}

Four such resimulations are compared with the \borg originals in \cref{fig:gadget_resims} where the present-day particle distributions in the Local Group region are plotted on top of each other. Although the two integration schemes do produce similar LG analogues, there are sometimes noticeable offsets between the halo positions of the Milky Way and Andromeda. In these cases, the corresponding velocities are biased oppositely, suggesting that this difference affects mostly the phase of the orbit and not the overall properties. We discuss the amplitude of these differences in more detail below.

\begin{figure}
    \centering
    \includegraphics[width=1\linewidth]{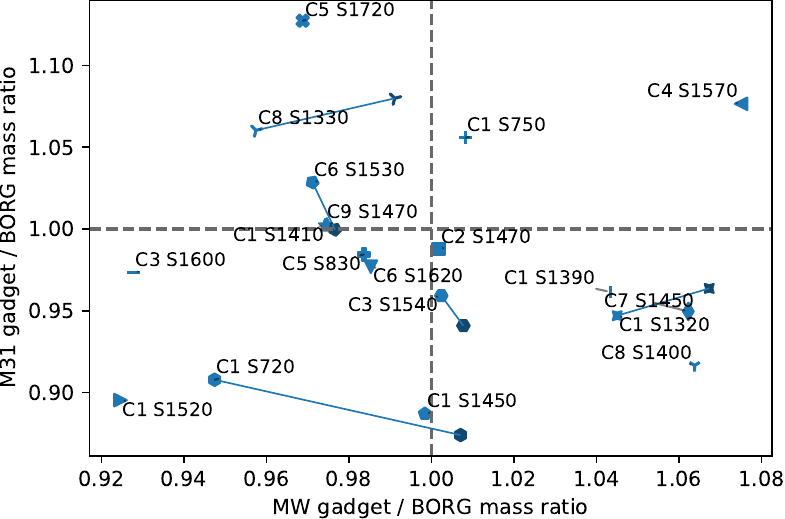}
    \caption{Filtered mass ratios of the Gadget resimulations compared to \borg (after applying the correction from \cref{app:masscorrection}). Each symbol indicates a different set of initial conditions. For five simulations, we also show the result with higher Gadget resolution (with particle masses $64\times$ smaller, softening $16\times$ smaller and power included in the initial conditions to $4\times$ smaller scale) in darker colours. Points referring to simulations with the same initial conditions but different resolutions are connected with a blue line.}
    \label{fig:gadget_resim_massratio}
\end{figure}
\begin{figure}
    \centering
    \includegraphics[width=1\linewidth]{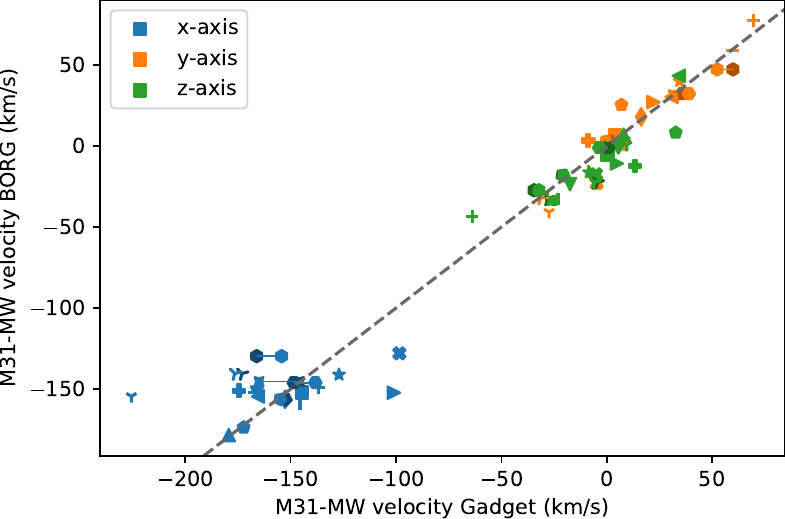}
    \caption{Peculiar velocity differences between the M31 and MW analogues in the Gadget resimulations plotted component by component against the same quantity for the \borg simulations. The x-axis is the Sun-M31 axis, and y and z are oriented orthogonally in directions of increasing R.A. and Dec. at the position of M31. As in \cref{fig:gadget_resim_massratio}, pairs  of symbols joined by lines refer to the five cases with an additional higher resolution Gadget resimulation.}
    \label{fig:gadget_resim_veldiff}
\end{figure}
\begin{figure}
    \centering
    \includegraphics[width=1\linewidth]{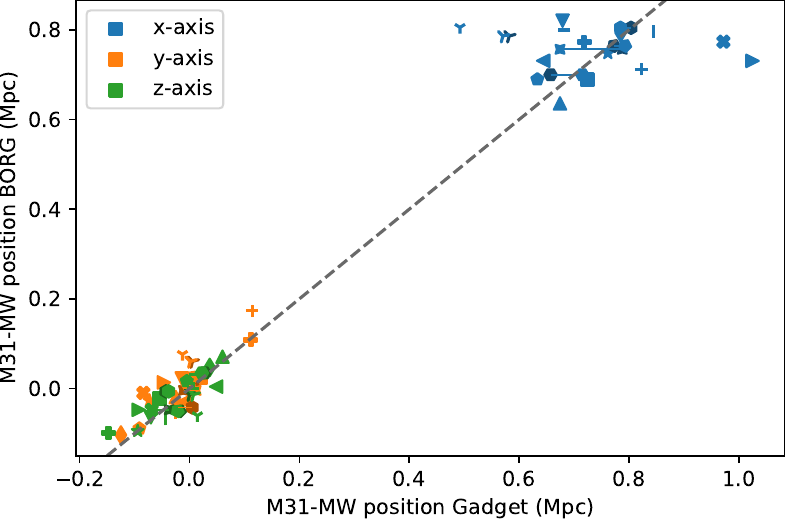}
    \caption{Position differences of the M31 and MW analogues in the Gadget resimulations plotted against the same difference in the \borg simulations. The x-axis is the Sun-M31 axis, and y and z are R.A. and Dec. directions at the position of M31. As in \cref{fig:gadget_resim_massratio}, pairs of symbols joined by lines refer to the five cases with an additional higher resolution Gadget resimulation.}
    \label{fig:gadget_resim_posdiff}
\end{figure}

The filtered masses of the BORG run and the Gadget run match quite well after correcting the \borg masses following the prescription of \cref{app:masscorrection}.
This can be seen in \cref{fig:gadget_resim_massratio}, where the corrected \SI{100}{kpc} filtered masses of the Milky Way and M31 in the \borg simulation are compared with the corresponding masses in the Gadget re-runs. 
To be able to compare masses properly, we filter based on the actual halo centres, as determined by a shrinking-spheres procedure \citep{powerInnerStructureLCDM2003} for both the Gadget and the \borg runs. 
The filtered mass ratios are generally close to one, with some small spread, which is, however, expected (about \SI{6.5}{\percent} is expected, see \cref{fig:masscorrection}).%

Positional offsets between the main haloes in the \borg-Gadget simulation pairs are typically at the level of 100~kpc and in a few realisations up to 300 kpc, as can be seen in \cref{fig:gadget_resim_posdiff}, where the three components of the MW-M31 position difference are shown.
The peculiar velocity difference between the two haloes also differs somewhat between \borg and Gadget, as can be seen in \cref{fig:gadget_resim_veldiff}, where the symbols are the same as in \cref{fig:gadget_resim_posdiff}. We note that cases with a large relative position offset also have a large relative velocity offset, and furthermore, these offsets are strongly correlated: where Gadget gives larger separations, it gives a smaller relative velocity, and vice versa. Forward (or backward) integration of the offset Gadget simulations into their relatively near future (or past) would likely bring them into substantially better agreement with the observational constraints. 

For five of the cases, we have also carried out simulations with substantially improved mass and spatial resolution. Specifically, for the HR particles these additional resimulations use a $64\times$ smaller particle mass and a $16\times$ smaller softening length; furthermore, they extend the power in the initial conditions to $4\times$ higher spatial frequency\footnote{Considerable care is needed to include this extra small-scale power consistently while retaining the amplitudes of the longer wavelength modes from the lower resolution simulation. We will describe this in detail in a future publication dedicated to such high-resolution resimulations.}. This may be expected to add further random offsets in the evolutionary phase of the MW-M31 orbit on top of those found for our main set of resimulations. (See \citet{sawalaSIBELIUSProjectPluribus2022} for further discussion of the influence of initial conditions on phase offsets in the orbits of LG analogues.) We tested that the effects of the increase in mass resolution and of the extra small-scale power on the masses, positions and velocities of the two main haloes are relatively minor, as may be seen in \cref{fig:gadget_resim_massratio,fig:gadget_resim_veldiff,fig:gadget_resim_posdiff}. 

\section{Discussion}
\label{sec:discussion}
In \cref{tab:previoussims}, we listed some major recent simulation studies of Local Group analogues in their observed cosmological context. These studies found their analogues by carrying out large numbers of \LCDM simulations constrained to reproduce nearby large-scale structures (smoothed over scales of $\SI{\sim5}{Mpc}$) and then selecting the few that, by chance, produced systems approximately matching the observed LG. Because we use the observed properties of the Local Group and its environment as explicit constraints in our \LCDM sampling, we match or surpass the requirements of the most precise previous programme, while including additional constraints, for example, requiring a match to the observed quiet Hubble flow in the environment of the Local Group out to 4~Mpc. Further constraints could easily be included, for example, on the spin orientation of the MW and M31 haloes, or on the properties of well-observed massive nearby groups such as M81 or Centaurus. Our current effective sample size, (\num{\sim27} quasi-independent realisations) is relatively small, but additional realisations can be generated simply by continuing the sampling chains as long as is needed. Although the resolution of these simulations is quite low, offsets when resimulating at higher resolution are relatively small.

\subsection{Comparison to previous Local Group simulations}
In comparison to the simulations listed in \cref{tab:previoussims}, our simulations generally enforce more precise constraints on MW and M31 properties like halo masses, separation, and velocity. The one exception is the radial velocity constraint of Sibelius-``Strict" \citep{sawalaSIBELIUSProjectPluribus2022}). They require a relative radial velocity in the range 99 to \SI{109}{km.s^{-1}}, whereas we adopt a Gaussian uncertainty of dispersion \SI{14}{km.s^{-1}}. 
We have also not placed any constraints on the presence of a massive satellite like the LMC or M33, which was done when refining the selection of LG candidates in e.g. ELVIS \citep{garrison-kimmelELVISExploringLocal2014}. 
For the Local Group's environment, we constrained the (quiet) local Hubble flow using 31 observed flow tracers. Only APOSTLE \citep{fattahiApostleProjectLocal2016} has a related constraint, which they put on the mean Hubble flow magnitude at \SI{2.5}{Mpc}. We have placed no constraints on structure more distant than \SI{4}{Mpc} \citep[unlike Clues, Hestia and Sibelius][]{carlesiConstrainedLocalUniversE2016a,libeskindHESTIAProjectSimulations2020,sawalaSIBELIUSProjectPluribus2022}. As a result, our simulations do not generally have a Virgo cluster analogue, although an appropriate large-scale tidal field at the Local Group's position is enforced by our detailed fitting of the local (perturbed) Hubble flow.
Finally, our current simulations are gravity-only/dark-matter-only; we intend to carry out higher resolution, full physics resimulations in future work.

\subsection{The mass distribution in and around the Local Group}
\label{sec:massdistributionlgdiscussion}
We use four classes of observations to constrain our analysis: dynamical tracers in the MW halo, dynamical tracers in the M31 halo, the relative 3D position and velocity of the MW/M31 pair, and the perturbed Hubble flow in the immediate environment of the Local Group. Much previous work has addressed each of these topics. Some of the authors of previous work directly or indirectly adopt a \LCDM prior, but most prefer more idealised and less specific model assumptions. As a result, it is not easy to compare masses obtained by different methods and in different regions to decide whether they are compatible. In contrast, our procedure randomly samples realisations subject to {\it all} the constraints simultaneously, leading to consistent estimates of the mass in all regions of the LG and its environment under a \LCDM prior. 

For technical reasons, we have expressed prior observational constraints on the masses of the individual MW and M31 haloes in terms of their mass profiles filtered with a Gaussian of dispersion \SI{100}{kpc}. 
Because of the limited resolution of our forward modelling, the mass on smaller scales is not well reproduced by our current scheme, and a bias correction is still needed even for 100~kpc-filtered masses.
We thus limit the quantitative comparison with earlier work to similar scales.
In their recent exhaustive review of the MW case \citet{wangMassOurMilky2020} convert all previous studies to estimates of the virial mass of the MW halo, finding results in the range 
$11.75<\log_{10}(M_{200c}/M_\odot)<12.42$ when all results are considered, or $11.89<\log_{10}(M_{200c}/M_\odot)<12.12$ using only recent results that make use of proper motion information from Gaia-DR2.
Different results in the first set are incompatible according to their published errors, which may be due to unknown systematics, whereas those in the second set are mutually compatible and appear to exclude $\log_{10}(M_{200c}/M_\odot)>12.2$ at about $2\sigma$. 
From our chains, we find $\log_{10}(M_{200c}/M_\odot) = \num{12.075(079)}$, which is similar to our adopted constraint on the  mass; our analysis of a limited set of observational data in \cref{app:observationalmasses} leads to $\log_{10}(M_{200c}/\si{\Msun}) = \num{12.053(081)}$.
For M31, the three outer halo mass estimates discussed in \cref{app:observationalmasses}, are consistent with each other within their quoted error bars, and combining them assuming an NFW-like profile gives a virial mass estimate $\log_{10}(M_{200c}/M_\odot) = 12.31\pm0.13$. For comparison, the result from our chains is  $\log_{10}(M_{200c}/M_\odot) = 12.331 \pm 0.092$. Thus, the posterior mass estimates are not pulled far from the direct observational constraints, derived and reported in \cref{app:observationalmasses}. This is remarkable as constraints from the other galaxy, from the Timing Argument and from the surrounding Hubble flow were not enforced in the estimates from \cref{app:observationalmasses}.

The separation and relative velocity of the MW/M31 pair were first used to estimate its total mass by \citet{kahnIntergalacticMatterGalaxy1959}.
Their Timing Argument idealised the system as a pair of point masses, and more recent work has often quoted masses based on a similar underlying model \citep[e.g.][the latter include the LMC as a third subdominant point mass, and treat the surrounding Hubble flow as perturbed only by the sum of these three masses]{vandermarelLGmass2008,penarrubiaTimingConstraintTotal2016}. Given the extended nature of the matter distribution in the Local Group, it is unclear what region the mass returned by such analyses should be considered to apply to. This difficulty can be avoided by adopting an explicit \LCDM prior. 
\citet{liMassesLocalGroup2008} searched a large \LCDM simulation and identified many isolated pairs of haloes comparable in individual halo mass, separation and relative velocity to the Local Group. 
They then used these pairs to calibrate the relation between the mass inferred from the Timing Argument and the sum of the virial masses of the two haloes, finding the two to be similar\footnote{\citet{sawalaTimelessTimingArgument2023a} show that this is only true for systems which are similar to the Local Group in that the sum of the two virial radii is similar to the separation of the haloes.} and to be proportional to each other with fairly small scatter, arriving at the final estimate $\log_{10}(\Sigma M_{200c}/M_\odot) = 12.72\pm 0.185$. 
We find with our analysis $\log_{10}(\Sigma M_{200c}/M_\odot) = 12.53\pm 0.07$, which is in tension at $1\sigma$ with the above past results. Although formally still consistent, this demonstrates that including the additional constraints on individual halo masses and the surrounding Hubble flow pulls the Timing Argument mass as derived from just properties of the individual haloes by \citet{liMassesLocalGroup2008} to substantially lower values. Note that our results have a much-reduced uncertainty. \citet{vandermarelM31VelocityVector2012} pointed out both the lower mean value and the smaller uncertainty found when Timing Argument results are combined with mass estimates for the individual haloes, but they did not consider data on the nearby Hubble flow, nor did they explicitly require simultaneous consistency with all observed data in a \LCDM cosmology; nevertheless, their final estimate of the total LG mass converts to $\log_{10}(\Sigma M_{200c}/M_\odot) = 12.42\pm 0.08$, slightly lower than our result here.

\citet{penarrubiaDynamicalModelLocal2014} use a catalogue of nearby galaxies very similar to that of \cref{fig:eddall} to estimate the total mass of the Local Group. In their model, all deviations from uniform undecelerated expansion are due entirely to the 
attraction of a point mass put at the LG barycentre.
Perhaps surprisingly, this model provides a reasonable fit to the observed velocities of galaxies lying between 0.8 and \SI{3}{Mpc} from the LG-barycentre, even though it assumes that all mass within \SI{3}{Mpc} is contained in the LG itself (see \cref{fig:eddall}). 
It leads to an estimated total LG mass, $\log_{10}(M_{LG}/M_\odot) = 12.36\pm 0.14$.
This is much smaller than the masses $\log_{10}(M(\SI{<0.8}{Mpc})/M_\odot) = 12.68\pm 0.05$ and $\log_{10}(M(\SI{<3.0}{Mpc})/M_\odot) = 12.96\pm 0.08$ derived from our analysis within the inner and outer radii of the nearby galaxy sample they used. 
These estimates are, respectively, factors of 2 and 4 larger than our estimate $\Sigma M_{200c}$ for the sum of the MW and M31  virial masses, showing that \LCDM analogues of the Local group typically have substantial additional mass in these larger regions.
Despite this, \cref{fig:hubblesim} shows that our \LCDM realisations fit the quiet local Hubble flow about as well as the model of \citet{penarrubiaDynamicalModelLocal2014}.
The latter must clearly fail at larger distances since already at \SI{3}{Mpc} material spread uniformly at the mean cosmic density would contribute about half of the enclosed mass (see \cref{fig:massprofiles}). 
On the other hand, the velocities of nearby galaxies are clearly not compatible with values of $\Sigma M_{200c}$ at the upper end of the range allowed by the $\Lambda$CDM-calibrated Timing Argument analysis of \citet{liMassesLocalGroup2008}, and so must contribute substantially to the strong downward pull noted in the last paragraph.

\subsection{Improvements and future directions}
In the future, we could consider many different avenues for improving our HMC methods. 
Firstly, a more accurate gravity solver could improve the accuracy of the reconstructions. This could be achieved in the first instance by higher resolution at the cost of increased computational cost, but there may be scope to develop more sophisticated approaches that still allow the adjoint gradient to be calculated.
To run all the chains presented in this paper, we used a total of about \num{800000} core-hours (logical, i.e. with hyperthreading/SMT, physical: \num{400000} ).
This could reduced by more careful tuning of the code, for instance, an improved method for load-balancing.
The resimulations at unchanged mass resolution presented in this paper are very cheap, requiring only about \num{25} core-hours per simulation, but a real resimulation program would, of course, substantially improve the mass resolution in addition to the spatial and time resolution, leading to increased computational cost (the simulations at $64\times$ improved mass resolution mentioned at the end of \cref{sec:resimulations} each took about \num{2000} core-hours).

Improvements could also be made to the sampling method. At the moment, the HMC mass matrix is diagonal; for instance, tuning the mass ratio between the LR and HR cells (this is currently unity) might be beneficial. 
Replacing the HMC altogether with a more efficient algorithm might also be possible, for example, the No-U-turn sampler might be worth exploring, which can, in some circumstances, outperform even a well-tuned HMC \citep{hoffmanNoUTurnSamplerAdaptively2011}.
The algorithm suggested by \citet{bayerFieldLevelInferenceMicrocanonical2023} might also be worth exploring but would require additional work to quantify its bias in our setup since it is not generally guaranteed to give unbiased results.

In terms of getting closer to observation, there are many more things to explore.
For instance, a goal might be to incorporate the Large Scale Structure reproduced in previous \borg applications. 
This could be done either by sampling fully or by optimising small-scale fields overlaid on fixed large-scale initial conditions predetermined by \borg. 
On smaller scales, one could envision extending the likelihood to include the LMC and M33.
However, their haloes currently overlap substantially with the Milky Way and Andromeda respectively, so careful thought would be needed to implement such constraints optimally. 
For instance, one might apply the constraints a few \si{Gyr} back in time, when the haloes had less overlap.

The current method for computing the velocity field at the locations of the galaxies is also not optimal: we consider the velocity field as a single-valued function, but in reality, dark matter motions are multi-valued, with multiple `streams' at many locations due to complex folding of the dark matter sheet in phase-space \citep{abelTracingDarkMatter2012}.
Additionally, one should also marginalise over distance uncertainties. Implementing a
a better method would require a substantial amount of work, especially considering the need for any estimator to be differentiable.

\section{Conclusion}
\label{sec:summary}
We have extended the \borg framework,  a Hamiltonian Monte Carlo scheme, by adding a zoom component to the gravity simulation and incorporating a robust Local Group likelihood based on observed mass tracers in the Milky Way and M31 haloes, on their relative position and velocity, and on the peculiar velocities of surrounding galaxies out to 4~Mpc. Using these constraints we have inferred a fair and representative sample of \LCDM initial conditions which evolve to produce a sizable sample of high-fidelity LG analogues.

Almost all these analogues live in a wall-like structure that is roughly aligned with the Supergalactic Plane
and have a MW-M31 orbit which is very nearly radial; the mean and {\it rms} scatter of the posterior tangential velocity distribution is $(v_y, v_z) = (\num{12.7(213)}, \num{-6.7(187)})~\si{km.s^{-1}}$, where $y$ and $z$ directions are pointed orthogonally to directions of increasing R.A. and Dec at the position of M31.  Requiring simultaneous consistency with all our dynamical constraints results in masses of $\log_{10}(M_{200c}/M_\odot) = 12.07\pm0.08$ and $12.33\pm0.10$ for the Milky Way and Andromeda, consistent with the injected estimates based on halo mass tracers alone. Our combined analysis estimates the sum of the two halo masses to be $\log_{10}(\Sigma M_{200c}/M_\odot)= 12.46\pm0.07$ and their ratio to be $M_{200c,MW}/M_{200c,M31}= 0.57\substack{+0.12 \\ -0.16}$.
For the spherically enclosed mass of the Local Group, we find $M(<\SI{1}{Mpc}) = \SI{5.20(67)e12}{\Msun}$ and $M(<\SI{2.5}{Mpc}) = \SI{7.77(127)e12}{\Msun}$.

We also find that in high-resolution resimulations of our initial conditions, we match halo masses for the two big galaxies within \SI{10}{\percent} and positions within \SI{100}{kpc} or so. 
Some direct applications of such simulations include investigating the spatial and dynamical distributions of the satellites of the Milky Way and Andromeda galaxies, and the range of predicted growth histories for the two galaxies in the context of \LCDM. 

\section*{Software}
During this work, we have made use of various software packages. This includes \borg  \citep{jascheBayesianPhysicalReconstruction2013,jaschePhysicalBayesianModelling2019}, \textsc{Gadget-4} \citep{springelSimulatingCosmicStructure2021}, \textsc{xtensor}\footnote{\url{https://github.com/xtensor-stack/xtensor}}, \textsc{onetbb}\footnote{\url{https://github.com/oneapi-src/oneTBB}}, \textsc{fftw} \citep{frigoDesignImplementationFFTW32005}, \textsc{openmpi} \citep{gabriel04:_open_mpi}, \textsc{mpich}\footnote{\url{https://www.mpich.org}}, \textsc{jax} \citep{jax2018github}, \textsc{mpi4py} \citep{dalcinMpi4pyStatusUpdate2021}, \textsc{mpi4jax} \citep{mpi4jax}, \textsc{xarray} \citep{hoyer2017xarray,hoyer_2023_10023467}, \textsc{numpy} \citep{harris2020array}, \textsc{scipy} \citep{2020SciPy-NMeth}, \textsc{matplotlib} \citep{Hunter:2007}, \textsc{h5py}\footnote{\url{https://www.h5py.org/}}, \textsc{astropy} \citep{astropycollaborationAstropyProjectBuilding2018,astropycollaborationAstropyProjectSustaining2022}, \textsc{r3d} \citep{powellR3dSoftwareFast2015}, \textsc{adjusttext} \citep{ilya_flyamer_2023_10016869}.

\section*{Acknowledgements}
This work has
been financially supported by a Spinoza Prize from NWO to AH. GL acknowledges support from the Simons Foundation "Learning the Universe" Collaboration, the CNES Euclid, and the International Emerging Action (IEA) "Manticore" from CNRS.  JJ gratefully acknowledges support from the Swedish Research Council (VR) under project 2020-05143, 'Deciphering the Dynamics of Cosmic Structure,' and from the Simons Collaboration on 'Learning the Universe.' Additionally, JJ appreciates the hospitality of the Aspen Center for Physics, supported by National Science Foundation grant PHY-1607611, with participation funding provided by the Simons Foundation.
The computations were enabled by resources provided by the Swedish National Infrastructure for Computing (SNIC) at the PDC Center for High Performance Computing, KTH Royal Institute of Technology, partially funded by the Swedish Research Council through grant agreement no. 2018-05973. SS was supported by the Göran Gustafsson Foundation for Research in Natural Sciences and Medicine.
EW thanks Tom Callingham, Akshara Viswanathan, Andrija Kosti\'c and Vincent Souveton for helpful discussions.
This work has been done in part within the Aquila Consortium (\url{https://www.aquila-consortium.org}).

\bibliographystyle{aa}
\bibliography{references_zot,phd} %

\appendix

\section{Milky Way and M31 mass estimates}
\label{app:observationalmasses}
In our framework, we require filtered masses, as defined by \cref{eq:filtmass}, together with their observational uncertainties. It is not straightforward to find good estimates for these quantities from the literature based on independent mass tracers across a wide range of radii. In most analyses of specific mass tracers corresponding enclosed masses are quoted, but because different analyses tend to be based on different models and assumptions, combining these results takes some care. We, therefore, present here a new analysis of a subset of the available mass tracers for the Milky Way and Andromeda in an attempt to obtain robust and conservative estimates of their filtered halo masses with a \LCDM prior.

\subsection{Milky Way}
\begin{figure}[tpb]
  \centering
  \includegraphics[width=\linewidth]{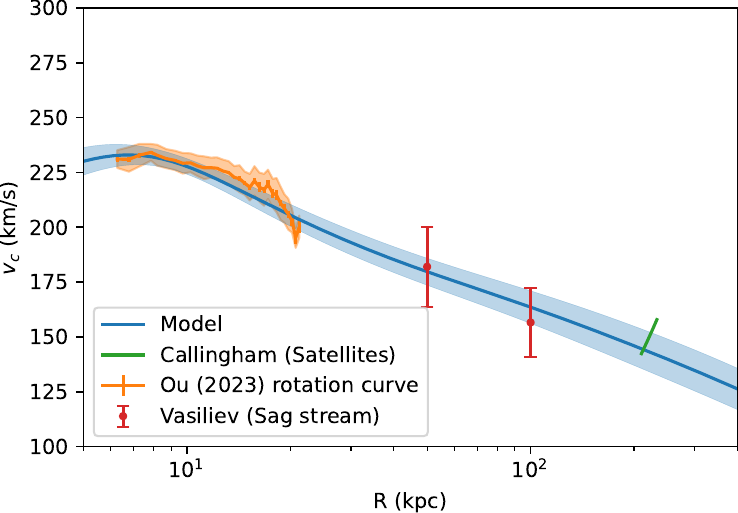}
  \caption{Our assumed circular velocity curve for the Milky Way, together with a band showing the $1\sigma$ scatter in the posterior distribution of \LCDM models constrained by the observational data points shown, with their adopted error bars, in orange, red and green.  The solid blue curve is the mean of the posterior circular velocity distribution at each radius. The error bar of the \citet{callinghamMassMilkyWay2019} estimate is diagonal since they estimated the virial mass and a lower mass results in a lower virial radius.}
  \label{fig:MWvc}
\end{figure}

For the Milky Way, there exist several joint analyses that constrain the mass profile using a heterogeneous set of mass tracers that go out to large radius \citep{mcmillanMassDistributionGravitational2017,karukesRobustEstimateMilky2020,wangMassOurMilky2020}. We choose a procedure, described below, which follows \citet{cautunMilkyWayTotal2020}, with a few minor alterations and additions to the data.
The mass model is the same as in \citet{cautunMilkyWayTotal2020} and consists of a bulge, a thin and thick disc, a circumgalactic medium, and a contracted NFW dark matter profile. We refer to the original paper for details. %
As constraints, we incorporate the following changes with respect to the fitting in \citet{cautunMilkyWayTotal2020}:
\begin{itemize}[nolistsep]
    \item We use the \citet{ouDarkMatterProfile2024} circular velocity curve. This is an update of the \citet{eilersCircularVelocityCurve2019} curve that was used by \citet{cautunMilkyWayTotal2020}. We only utilise the data points within \SI{<21.5}{kpc}, because the (systematic) errors become very large at larger radii.
    Since systematic errors dominate statistical errors in this analysis, we add to the statistical errors a covariance matrix that is modelled by a Gaussian Process \citep[see also][]{omanOverlookedSourceUncertainty2024}.
    We assume a rational quadratic covariance function; the covariance due to systematic errors between data points $i$ and $j$ is assumed to be $C(d_{ij}) = A^2\qty(1+d_{ij}^2/2\alpha k^2)^{-\alpha}$ where the width $k$ and power-law index $\alpha$ are allowed to vary, and where the amplitude $A$ is set to \SI{6}{km.s^{-1}}, which is the amplitude that \citet{ouDarkMatterProfile2024} estimate as the magnitude of their systematic error. For the distance metric $d_{ij}$ between data points $i$ and $j$, we use the difference in logarithm of Galactocentric radii $r_i$ and $r_j$, i.e. $d_{ij}=|\log{r_i/r_j}|$. We verified that the results obtained are robust against reasonable variation of covariance function and distance metric. One can see that with this treatment, the uncertainty of the final fit (\cref{fig:MWvc}) reflects well the systematic uncertainty in the inner region, something that is not achieved when simply adding the systematic uncertainty in quadrature with the statistical uncertainty.
    \item We add mass estimates from Sagittarius stream modelling by \citet{vasilievTangoThreeSagittarius2021}.
    Although they have a different (more flexible) dark matter profile, the enclosed masses are given too.
    They quote two spherically enclosed masses (the constraint is put on the integrated density within this sphere): $M(\SI{50}{kpc})=\SI{3.85(10)e11}{\Msun}$ and $M(\SI{100}{kpc})=\SI{5.7(3)e11}{\Msun}$.
    However, their model is triaxial and includes an LMC.
    Since we limit ourselves to a simple spherical model, we opt for an error of \SI{20}{\percent} to be able to generalise the masses to our spherical model.
\end{itemize}
The fit and the model velocity profiles are shown in \cref{fig:MWvc}. For the dark matter profile, we have sampled the NFW parameters of the dark-matter-only analogue, that, after using the recipe from \citet{cautunMilkyWayTotal2020} resulted in the contracted NFW profile we have used in our mass model. 
Since our simulations are dark-matter-only, this sample of implied dark-matter-only analogues is ideal for constraining our simulations.
We scale up these sample profiles by a factor $(1-f_\text{bar})^{-1}$ to account for the baryon fraction, $f_\text{bar}=0.157$.  For each of these scaled NFW profiles, we then compute the implied \SI{100}{kpc} filtered mass following \cref{eq:filtmass} as $M_\text{filt}(\SI{100}{kpc}) = \int_0^\infty {\rho_\text{NFW}(r)\exp(-r^2/2\sigma^2)4\pi r^2\dd{r}}$. 
We find $\ln(M_\text{filt}(\SI{100}{kpc})) = \ln(\SI{1.085e12}{\Msun}) \pm 0.1291$, the value listed in \cref{tab:massestimatesadopted} as our constraint, where we also include an additional \SI{5}{\percent} in quadrature to account for the halo-to-halo scatter that \citet{cautunMilkyWayTotal2020} found.
\subsection{M31}
For M31, we carry out an analysis similar to that for the Milky Way, also with a contracted halo.
\begin{itemize}[noitemsep]
    \item We use a stellar + gas mass model from \citet{tammStellarMassMap2012}, derived purely from stellar population modelling. This model consists of a nucleus, a bulge, a stellar disc, a star-forming young disc and a stellar halo. Since different choices in the stellar population models allow for up to \SI{50}{\percent} extra stellar mass (this freedom mostly stems from the choice of stellar initial mass function) \citep{tammStellarMassMap2012}, we include an unknown multiplicative factor $A$, i.e. $\rho=A\rho_\text{tamm, min}$, where $A\sim \mathcal{U}(1,1.5)$ and $\rho_\text{tamm, min}$ is their minimum stellar-mass model.
    \item The NFW dark matter profile is contracted, following the procedure of \citet{cautunMilkyWayTotal2020}
\end{itemize}
We derive constraints as follows:
\begin{itemize}[noitemsep]
    \item We consider the rotation curves derived by \citet{cheminKinematicsDynamicsMessier2009,corbelliWidefieldMosaicMessier2010,carignanExtendedRotationCurve2006}.
    There is some scatter within the literature on measurements of the rotation curve. 
    Since we are interested only in the total mass within our rather large Gaussian filter, it is not useful to model the full internal structure of M31. We therefore use only one data point as a constraint, which we choose to put at $r=\SI{35}{kpc}$, since the datasets do not go beyond this.
    In order to smooth out small-scale structure, we average velocity measurements within a \SI{5}{kpc} window centred at \SI{35}{kpc} for each dataset. Although reported uncertainties are typically small, they often do not include the systematic difference between the two sides of the galaxy, which we do not attempt to model and hence must include as an uncertainty. However, the errors reported by \citet{corbelliWidefieldMosaicMessier2010} do include these, so we adopt their average uncertainty within the window. This results in our chosen constraint, $v_c(\SI{35}{kpc}) = \SI{239(25)}{km.s^{-1}}.$
    \item \citet{fardalInferringAndromedaGalaxy2013} used a Bayesian sampling method based on N-body modelling of the giant stellar stream to estimate M31's halo mass. Since any quoted $M_{200c}$ is model-dependent, we convert this to an enclosed-mass estimate within their fitted apocentre, $M(r_\text{apo}=\SI{55.5}{kpc}) = \SI{5.92(3)e11}{\Msun}$, which we expect to be robust. This uncertainty is an underestimate, however, because the halo scale radius, disc and bulge mass were not taken as free parameters; rather, \citet{fardalInferringAndromedaGalaxy2013} determined them as a function of halo mass, such that in total the model roughly matches the rotation curve. This is done by assuming that the parameters lie exactly on the maximum-likelihood curve of the fit by \citet{geehanInvestigatingAndromedaStream2006}. To get the true uncertainty we should therefore propagate the uncertainty of the \citet{geehanInvestigatingAndromedaStream2006} model and add it to our uncertainty estimate. Although doing this exactly is impractical, we can get a conservative estimate on the error of $M(r_\text{apo})$ by assuming that it is only known as well the the rotation curve itself. Hence, we use the same relative error on this data point as on the rotation curve point discussed above, leading to the constraint $M(<\SI{55.5}{kpc}) = \SI{5.92(124)e11}{\Msun}$.
    \item \citet{veljanoskiOuterHaloGlobular2014} use outer-halo globular clusters as mass tracers. They estimate an enclosed mass of $M(<\SI{200}{kpc}) = \SI{1.2(2)e12}{\Msun}$ for an NFW-like potential. 
    \item \citet{tollerudSPLASHSurveySpectroscopy2012} estimate a halo mass from satellite galaxy data. They utilise an estimator of the form $M(<r) = C(r) \sigma^2 r/G$, tuning $C(r)$ to match simulation data. This leads to an estimated enclosed mass  within the satellite median distance of $M(<\SI{139}{kpc}) = \SI{8.0(39)e11}{\Msun}$.
    \item \citet{patelEvidenceMassiveAndromeda2023} compare satellite dynamics directly with simulations to infer a virial mass (defined as the mass within a sphere of mean density $18\pi^2\rho_\text{crit}$) of $M_\text{vir} = 
    \SI[parse-numbers=false]{2.85\substack{+1.47 \\ -0.77}\times10^{12}}{\Msun}$. We convert this to a log-normal constraint on the virial mass, setting the standard deviation to be $\sigma_{\ln{M}} = (\ln(2.85+1.47) - \ln(2.85-0.77)) / 2$.
    \end{itemize}
\begin{figure}[tpb]
  \centering
  \includegraphics[width=\linewidth]{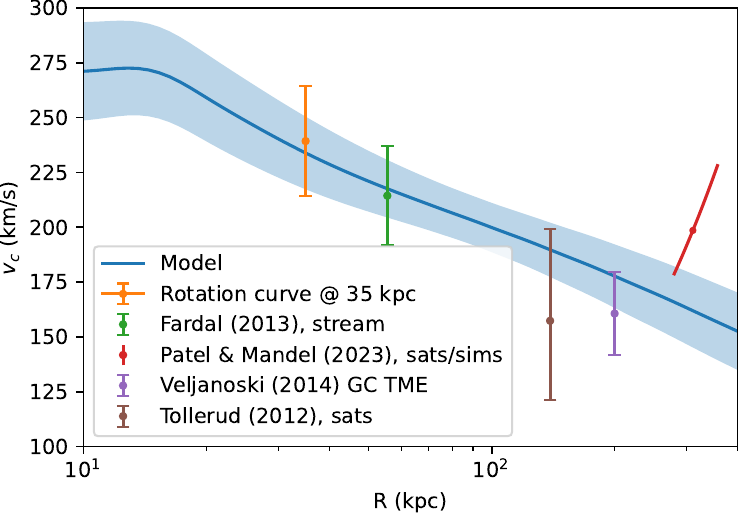}
  \caption{Our assumed \LCDM circular velocity curve for M31, together with a band showing the $1\sigma$  scatter in allowed models. The data points used for the fitting are also shown.}
  \label{fig:m31vc}
\end{figure}%
The circular velocity curve of our model is shown in \cref{fig:m31vc}. The final constraint on the filtered mass, listed in \cref{tab:massestimatesadopted}, is derived completely analogously to the one for the Milky Way described in the previous section. %
\section{Positions and velocities of the Milky Way and M31}
\label{app:coordinates}
\subsection{Milky Way}
For the position and velocity of the Sun with respect to the Milky Way, we use the default parameters of Astropy \citep{astropycollaborationAstropyProjectBuilding2018,astropycollaborationAstropyProjectSustaining2022}, which use measurements of the position and motion of the Sagittarius A* by \citet{reidProperMotionSagittarius2004,gravitycollaborationDetectionGravitationalRedshift2018,drimmelSolarVelocity2018}.
The height above the plane is from \citet{bennettVerticalWavesSolar2019}. 
Specifically, this is a solar Galactocentric velocity of $\vb*{v}_\odot = (12.9, 245.6, 7.78)\,\si{km.s^{-1}}$, with a Galactocentric distance of \SI{8.122}{kpc}, and a disk-height of $z_\odot=\SI{20.6}{pc}$.
The measurement uncertainties are of the order of a few \si{km.s^{-1}}, and hence we neglect them compared to the other sources of uncertainty in our analysis. In particular, we include an isotropic uncertainty of \SI{10}{km.s^{-1}} in the offset from the Solar frame of reference to that of the barycentre of the MW's filtered halo mass (see \cref{sec:veloffsetframe}).

\subsection{Andromeda}
The position on the sky of M31 is $(\alpha,\delta)=(00^h42^m44.33^s,  41^d16^m07.5^s)$ \citep{kafleNeedSpeedEscape2018}, and we take its distance modulus to be $\mu=\num{24.46(10)}$ \citep{degrijsClusteringLocalGroup2014}, which corresponds to a distance of $d_\text{M31}=\SI{780(46)}{kpc}$.

The heliocentric radial velocity of M31 is $v = \SI{-301(1)}{km.s^{-1}}$ \citep{watkinsCensusOrbitalProperties2013}. For the proper motions, we use the recent measurement by \citet{salomonProperMotionAndromeda2021} based on Gaia EDR3 data, specifically $(\mu_\alpha\cos{\delta},\mu_\delta)=(\SI{48.9(105)}{\micro as/yr}, \SI{-36.9(81)}{\micro as/yr})$.
We use this in preference to the HST proper motion by \citet{sohnM31VelocityVector2012,vandermarelM31VelocityVector2012}, because the HST measurement is based on only a small part of the galaxy (because HST has a small field of view), and therefore, for the global motion, the result is quite sensitive to modelling choices for the internal dynamics \citep{vandermarelM31VelocityVector2012}. The resulting values for our constraints are quoted in \cref{tab:massestimatesadopted}. %

\section{Low-resolution mass bias correction}
\label{app:masscorrection}

The integrator in our Hamiltonian Monte Carlo sampler requires a full particle-mesh simulation to be run at each step. To build a sufficiently large MCMC sample size, we require very fast, and therefore low-resolution simulations. To give a practical number, each independent sample needs about 500 HMC rejection/acceptance steps. Each step requires about 200 leapfrog steps. Thus, for each independent sample, we need to run 100~000 evolutionary simulations.
Unfortunately at low resolution, the inner regions of our haloes are not well-resolved and so are less concentrated than they should be.
This was already noted in the literature as a systematic effect  \citep[e.g.][]{erraniAsymptoticTidalRemnants2021} and as such we may account for it relatively accurately by an appropriate correction. 

For this purpose, we have run two unconstrained simulations with the same initial conditions. One has the same spatial, temporal, and mass resolution as the HR region in our inference set-up, while the other is a high-resolution Gadget simulation from the same initial conditions.
Their properties are listed in \cref{tab:simplesim}.
We then create a density field (with a mesh size of $\Delta x = \SI{26}{kpc}$), and find all local maxima of the \SI{100}{kpc} filtered mass field in the two simulations. 
Each maximum corresponds to a halo, so we obtain a catalogue of halo masses\footnote{When computing the filtered masses, we derive them at the cell centres, instead of at the halo centres. This leads to small offsets between the filter and halo centre, which will bias the masses slightly low, but since the offset is always less than $\Delta x/2=\SI{13}{kpc}$, we found it to be negligible.} and positions.
We then select the haloes with filtered mass \SI{>2.5e11}{\MSun}.
We lastly require isolation (no halo \SI{>1e11}{\MSun} within \SI{2}{Mpc}) to filter out close pairs, which might not have well-defined masses.
We cross-match the two catalogues to get a relationship between low-resolution and high-resolution haloes, requiring a position difference of less than \SI{200}{kpc} when matching. 

\begin{figure}[tpb]
    \centering
    \includegraphics[width=\linewidth]{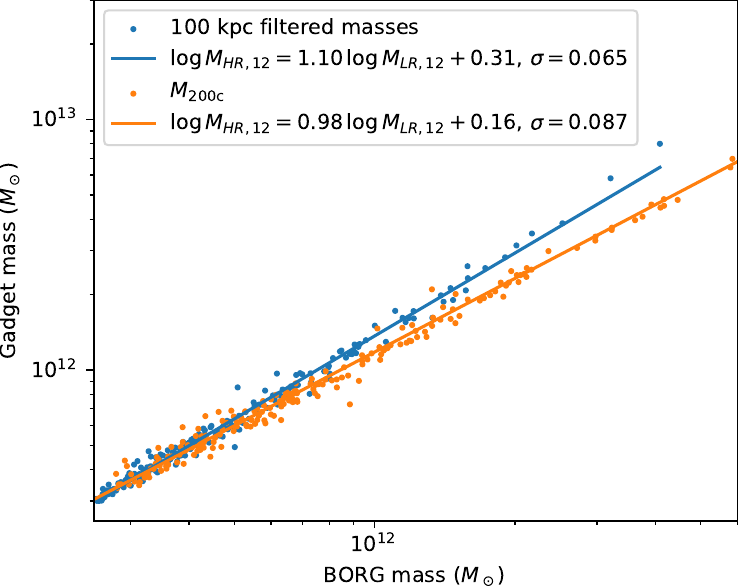}
    \caption{Halo masses in the high-resolution Gadget simulation compared to those in the low-resolution BORG simulation. $M_{HR,12}$ is the 100~kpc filtered mass (blue) or $M_{200c}$ value (orange) in the Gadget resimulation in units of \SI{1e12}{\Msun}, while $M_{LR,12}$ indicates the corresponding quantities for the BORG simulation. $\sigma$ indicates the scatter around the power-law fits to the relations given in the figure legend.}
    \label{fig:masscorrection}
\end{figure}

The resulting fits are shown in \cref{fig:masscorrection}.
Although there is some freedom in the choice of selection, isolation requirement, and cross-match precision, the results are robust against reasonable variation; the simple power-law fits given in the figure work quite well and are used in our main analysis.
\begin{table}
  \begin{center}
    \begin{tabular}[c]{lll}
      \toprule
      & \borg & Gadget \\
      \midrule
      Particle mass & \SI{1.5e8}{\Msun} & \SI{1.5e8}{\Msun}\\
      Box size & \SI{40}{Mpc} & \SI{40}{Mpc} \\
      Resolution & Cell size: \SI{78.125}{kpc} & Softening: \SI{10}{kpc/h}\\
      Timestepping & 40 steps & \num{>1000}, adaptive \\
      \bottomrule
    \end{tabular}
  \end{center}
  \caption{The simulation settings used for the halo mass bias correction.}
  \label{tab:simplesim}
\end{table}

\section{Marginalisation of the velocity frame}
\label{app:marginalisation}
Our velocity likelihoods consist of the 31 Gaussian distributions from the external galaxies and three Gaussian distributions for the three velocity components of M31. This can be written as a single multivariate Gaussian, $\vb*v_\text{mod} \sim \mathcal{N}(\vb*v_\text{obs}, \vb*\Sigma_\text{obs})$, where the components of the $\vb*v$ vectors are either the radial velocities of external galaxies, or one of the 3 components of the M31 velocity vector. The covariance matrix $\vb*\Sigma_\text{obs}$ is diagonal. 
As described in \cref{sec:veloffsetframe}, we wish to marginalise over a latent parameter, the velocity frame offset $\vb*v_\text{off}$ between the true Galactocentric reference frame, and the simulation-derived Galactocentric frame (which is defined such that the Milky Way's \SI{100}{kpc} filtered velocity is zero). In practice, we allow for a dispersion of $\sigma_\text{off}=\SI{10}{km.s^{-1}}$, because this is the standard deviation of the difference between \SI{100}{kpc} filtered velocity and the \SI{5}{kpc} filtered velocity of Gadget resimulations of our Milky Way of an earlier version of the chain that did not incorporate this additional uncertainty.

If $Z=X+Y$ where $X\sim \mathcal{N}(\vb*\mu_X, \vb*\Sigma_X)$ and $Y\sim \mathcal{N}(\vb*\mu_Y, \vb*\Sigma_Y)$, then $Z\sim \mathcal{N}(\vb*\mu_X + \vb*\mu_Y, \vb*\Sigma_X + \vb*\Sigma_Y$).
Moreover, if a scalar random variable $R \sim \mathcal{N}(\mu, \sigma^2)$ and a vector random variable $\vb*V\sim\mathcal{N}(\vb*\mu_V, \vb*\Sigma_V)$, are combined into a vector random variable $\vb*Y = \vb*V + R\vb*\alpha$, where $\vb*\alpha$ is a constant vector, then $\vb*Y$ will follow $\vb*Y\sim \mathcal{N}(\vb*\mu_{V} + \vb*{\alpha} \mu, \vb*\Sigma_V + \sigma^2 (\vb*\alpha \vb*\alpha^T))$.

Assume that there is an offset $\vb*v_\text{off}$ between the assumed heliocentric reference frame in the simulation and the true one and that each component has $v_\text{off,k}\sim \mathcal{N}(0, \sigma_\text{off}^2)$. The corrected radial velocity will be $v_\text{mod,i,corr} = v_\text{mod,i} + \vb*v_\text{off} \cdot \vu*r_i = v_\text{mod,i} + v_\text{off,x} \cdot \frac{x}{r} + v_\text{off,y} \cdot \frac{y}{r} + v_\text{off,z} \cdot \frac{z}{r}$.
In the case of M31, where we also have constraints on the tangential velocity, the corrected 3d-velocity would be $\vb*v_\text{M31,corr} = \vb*v_\text{M31} + \vb*v_\text{off}$. Now, if we define \begin{equation*}\alpha^{(i)}_k = 
    \hat{r}^{(i)}_k, 
\end{equation*}
where $\vu*{r}^{(i)}$ is the unit vector in the direction of the measured velocity giving rise to velocity constraint $i$, we can write
\begin{equation*}
    \vb*v_\text{mod,corr} = \vb*v_\text{mod} + \sum_{k=x,y,z}\vb*\alpha_k v_\text{off,k}
\end{equation*}
So, making use of the above, we have that 
\begin{equation*}
    \vb*v_\text{mod,corr} \sim \mathcal{N}(\vb*v_\text{obs}, \vb*\Sigma_\text{obs} + \sigma^2 (\vb*\alpha_x\vb*\alpha_x^T + \vb*\alpha_y\vb*\alpha_y^T + \vb*\alpha_z\vb*\alpha_z^T))
\end{equation*}

\label{LastPage}
\end{document}